# A physics-based framework for constraining near-surface tornadic winds from windborne-debris endpoints

Guangzhao Chen[a]*, Franklin T. Lombardo[b]

[a] Department of Civil, Environmental and Construction Engineering, Texas Tech University, TX, Guangzhao.Chen@ttu.edu

[b] Civil & Environmental Engineering Department, University of Illinois Urbana-Champaign, Urbana, IL, USA, lombaf@illinois.edu

**ABSTRACT:**

Near-surface tornado winds are seldom observed directly, whereas post-event imagery may retain both a debris source and the landing locations of windborne objects. This paper develops a physics-based inverse framework that uses a static, unpaired cloud of roof-panel endpoints to constrain a translating parametric vortex. The forward model combines probabilistic panel release, six-degree-of-freedom plate dynamics, orientation-dependent aerodynamic coefficients, and the Baker-Sterling vortex. Candidate fields are screened by comparing simulated and observed landing sets. A proof-of-concept application to the 2011 Joplin, Missouri tornado uses 17 mapped endpoints and 82 candidate panel-release locations. An executed finite-grid search using two release formulations and a symmetric point-set objective places simulated panels within the observed endpoint region, but it does not establish a unique wind field. The method therefore demonstrates a possible forensic constraint on near-surface winds; quantitative wind-field estimates remain conditional on release modeling, aerodynamic validation, time-step convergence, and reproducible parameter search.

## 1 INTRODUCTION

A tornado is an extreme and complex wind event that is composed of a violently rotational wind field and a persistently translational wind field, and it causes nearly one-fifth of all natural hazard fatalities based on 10-year average data in the United States. (NOAA, 2025). Near-surface tornado flow is difficult to measure directly. Radar observations often sample above the heights at which residential damage occurs, while post-event wind estimates depend on the availability and capacity of suitable damage indicators. Engineering vortex models provide a complementary description of the flow. Rankine and Burgers-Rott formulations represent idealized velocity structure (Burgers, 1948; Rankine, 1882; Rott, 1958), laboratory simulators characterize near-surface behavior (Refan and Hangan, 2018), and the Baker-Sterling model supplies a compact three-component field for repeated engineering calculations (Baker and Sterling, 2017). Chen and Lombardo (2019), for example, inferred parameters of the 2011 Joplin tornado from mapped tree-fall directions. Each method is conditioned on a different observable and a different vortex representation.

Windborne debris provides another post-event record. Classical studies established the flight mechanics of compact, rod-like, and plate-like objects in uniform winds (Tachikawa, 1983, 1988; Holmes et al., 2006; Lin et al., 2006, 2007), and later work extended trajectory calculations to three-dimensional and tornado-like flows (Huo et al., 2020; Abdelhady et al., 2021). A large volume of research encompasses the aerodynamic properties of the debris and flight trajectory characterization in wind events (Baker and Sterling, 2017; J.D. Holmes et al., 2006; Lin et al., 2007; Jafari and Sarkar, 2020). By simulating debris impacts and analysing the resulting damage patterns, researchers can develop a more comprehensive understanding of the relationship between the debris characteristics, and structure vulnerability in straight-line winds (Haan et al., 2008). Considering the debris trajectories in tornadic wind fields, Huo in 2020 apply extensive large eddy simulations for replicating tornado-like vortex and provide the trajectories of sphere debris (Huo et al., 2020). Zhao et al. (2021) provided a comprehensive overview of previous analytical and experimental methodologies used to investigate the dynamics of wind-borne debris in both straight-line winds and tornadic wind fields, offering a detailed synthesis of existing knowledge in the field. (Miller et al., 2024) introduced a Monte Carlo simulation based method for estimating wind speed by the moving track of large compact debris (i.e. Vehicles, construction equipment). Hence, the debris trajectories will contain information on tornadic wind fields, and it is possible to consider the windborne debris trajectories as evidence for reconstructing the near-surface tornadic field. Most of this literature addresses the forward problem: a prescribed wind field produces trajectories and impacts. Miller et al. (2024) inverted observed tracks of large compact objects to estimate wind speed, but track-based inference requires temporal information. Aerial damage surveys more often preserve a source building and a static cloud of endpoints without retaining trajectories or source-to-endpoint identities.

The question addressed here is whether an unpaired endpoint cloud can constrain a plausible near-surface wind field when release and flight are represented physically. The framework links a roof-panel release screen, a three-dimensional six-degree-of-freedom (6-DOF) plate model, orientation-dependent aerodynamic coefficients, and a translating Baker-Sterling vortex. A point-set objective ranks candidate fields without assuming one-to-one identity. The contribution is the inverse integration of these elements; no priority claim is made for 6-DOF debris dynamics. Sections 2-5 define the forward calculation. Section 6 presents the Joplin example and distinguishes the saved legacy calculation from the revised objective required for a reproducible reconstruction. Section 7 discusses uncertainty and identifiability.

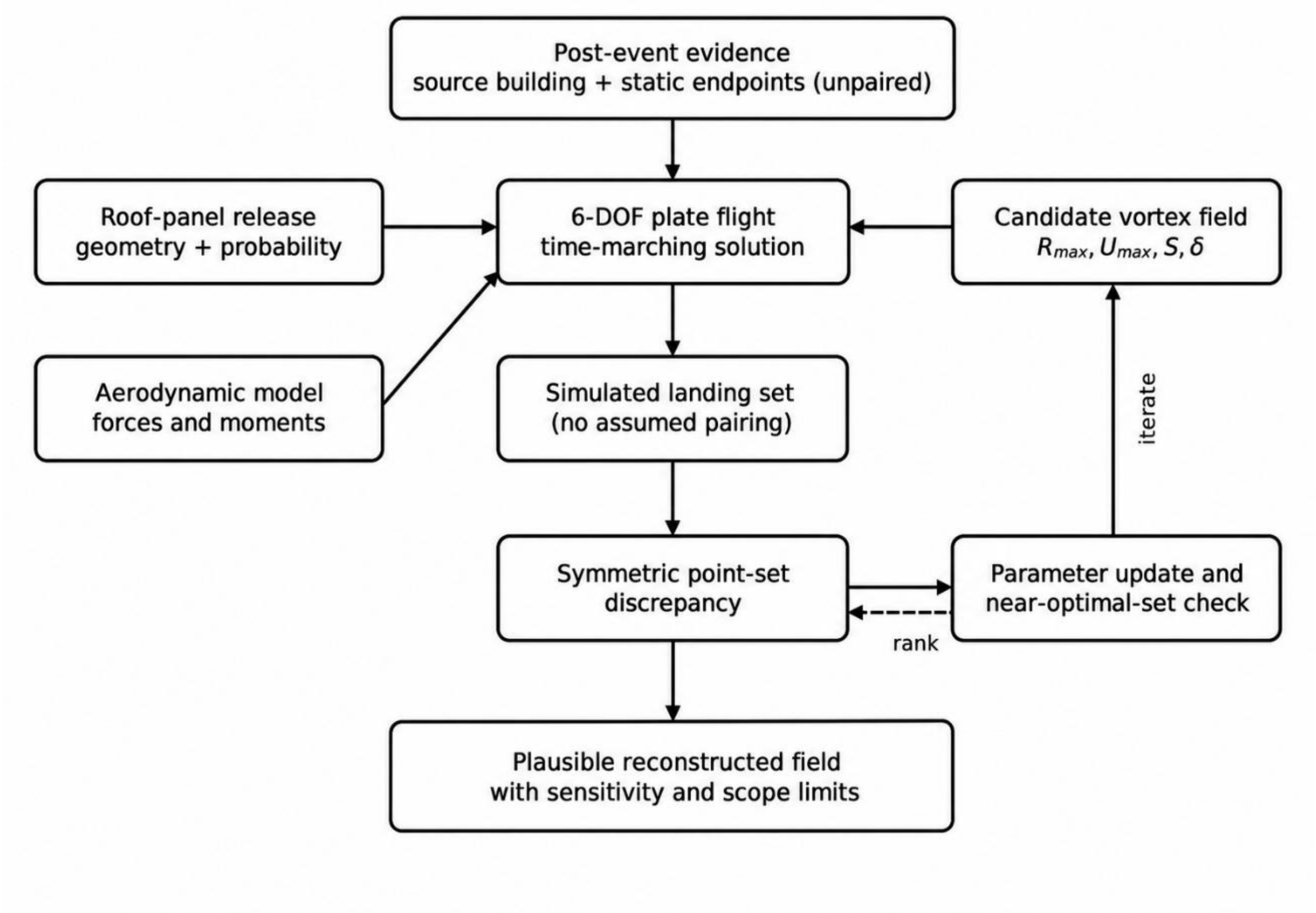

**Figure 1.** Forward-inverse workflow for using a known debris source and static, unpaired endpoints to screen candidate near-surface wind fields.

## 2 DEBRIS EQUATION OF MOTION (EOM)

The debris equations of motion map a local wind vector to translation and rotation of an object with specified geometry, mass, and inertia. Shape determines which aerodynamic forces and moments are active; size and density set the response time and the ratio of gravity to aerodynamic loading. Following Wills et al. (2002), debris is grouped as compact, plate-like, or rod-like (Figure 2). The reduced models below provide benchmarks, whereas the Joplin application uses the full 6-DOF plate formulation.

Early tornado-missile models treated drag on compact objects (McDonald, 1976). Tachikawa (1983, 1988) formulated the two-dimensional flight of flat plates, and subsequent studies added three-dimensional forces, moments, and numerical integration (Twisdale et al., 1979; Holmes et al., 2006; Baker, 2007; Richards et al., 2008). This progression leads from a transparent reduced-order benchmark to the body-fixed rigid-body equations needed for a freely rotating plate in a spatially varying vortex.

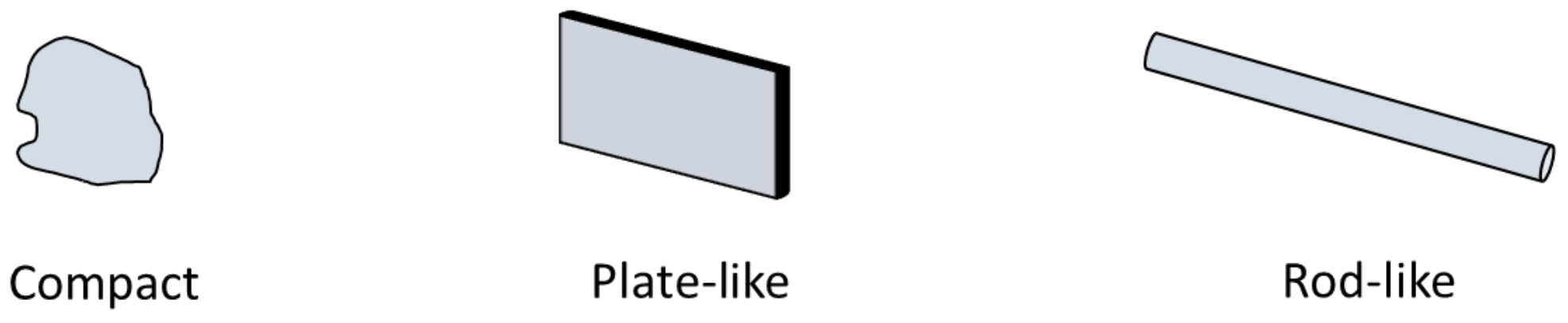

**Figure 2.** Geometric classes commonly used in windborne-debris modeling: compact, plate-like, and rod-like objects.

2.1 *Two-dimensional three-degree-of-freedom (3-DOF) general debris EOM under uniform flow*

As the foundational work, Tachikawa (1983) described plate debris moving two-dimensionally in an X-Z plane, which is also the debris principal axes coordinates, under a uniform horizontal flow $U$ illustrated in Figure 3. As for the debris, $a$ is the thickness, $b$ is the vertical length of the plate debris, and $c$ is the horizontal length. The debris is under the aerodynamic drag, lift, and moment denoted as Eq (1)-(3), respectively. In these equations, $\rho_a$ is the air density and is considered as a constant value (e.g., as 1.225 kg/$m^3$ at standard conditions), $A_c$ is the cross-section area perpendicular to the flow direction (e.g., as $A_c = bc$ in this case), and $L$ is the rotation radius (e.g., L=$b$ in this case). Then, the force analysis diagram in the global Cartesian coordinates is shown in Figure 4 considering the aerodynamic drag force $F_D$, aerodynamic lift force $F_L$, aerodynamic moment $F_M$, and gravity. As a result, the EOM of this debris is established as Eq. (4)-(6) (J. D. Holmes et al., 2006; Tachikawa, 1983, 1988). In this EOM, $C_D$, $C_L$ and $C_M$ are the drag, lift, and moment coefficients; $\rho_m$ is the density of the debris, and $I$ is the moment of inertia. Then, this debris is restricted to three degrees of freedom as the translation motion along X axis with speed $U_m$, Z axis with $V_m$, and rotation around Y axis with speed $\omega_m$.

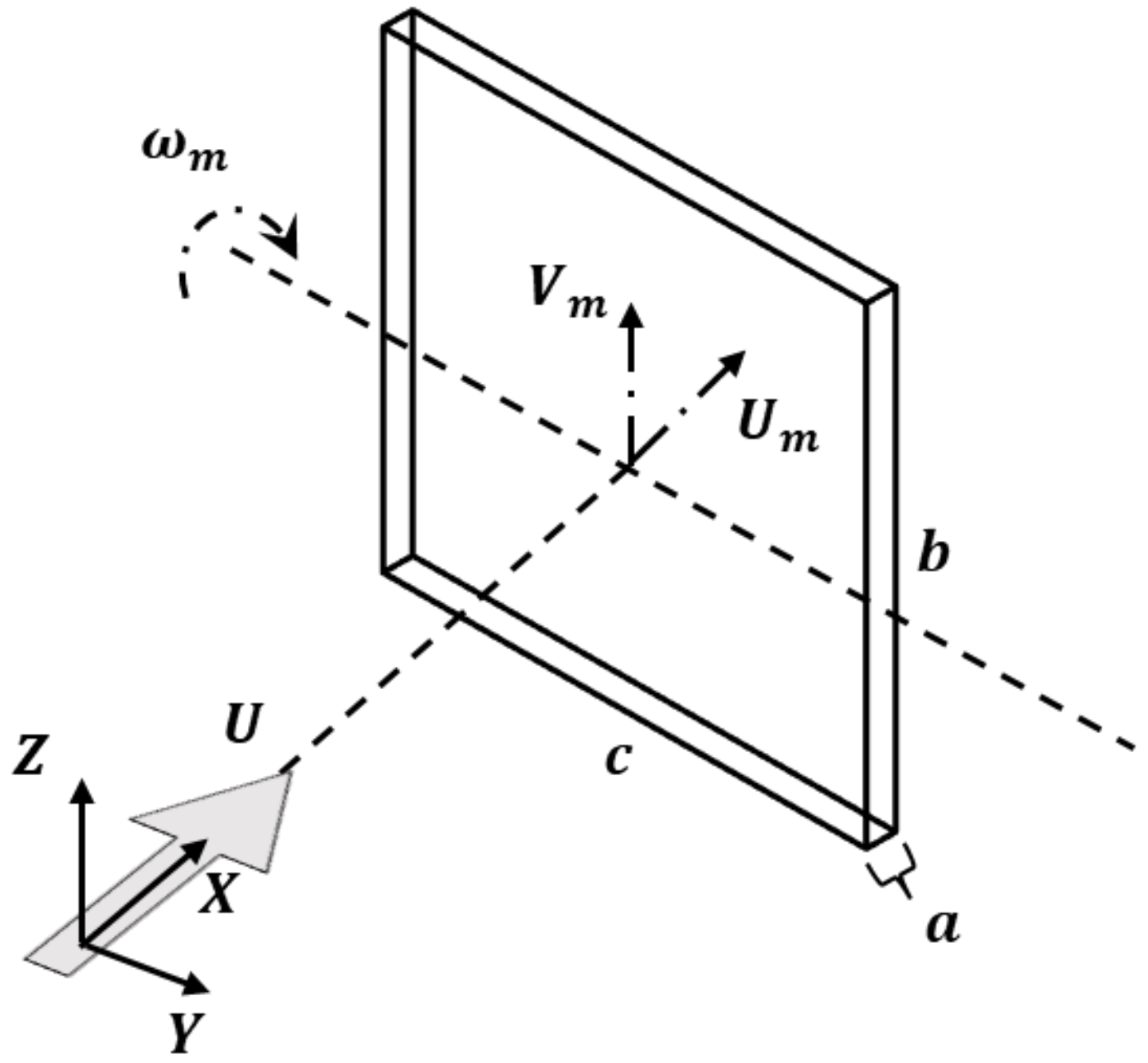

**Figure 3.** Plate in uniform flow with three degrees of freedom in the body-fixed principal-axis frame (orthogonal principal axes coordinates).

$$F_D = \frac{\rho_a}{2} A_c C_D \left[(U - U_m)^2 + {V_m}^2\right] \qquad \textit{(Eq.1)}$$

$$F_L = \frac{\rho_a}{2} A_c C_L \left[(U - U_m)^2 + {V_m}^2\right] \qquad \textit{(Eq.2)}$$

$$F_M = \frac{\rho_a}{2} A_c L C_M\left[(U - U_m)^2 + {V_m}^2\right] \quad \text{(Eq.3)}$$

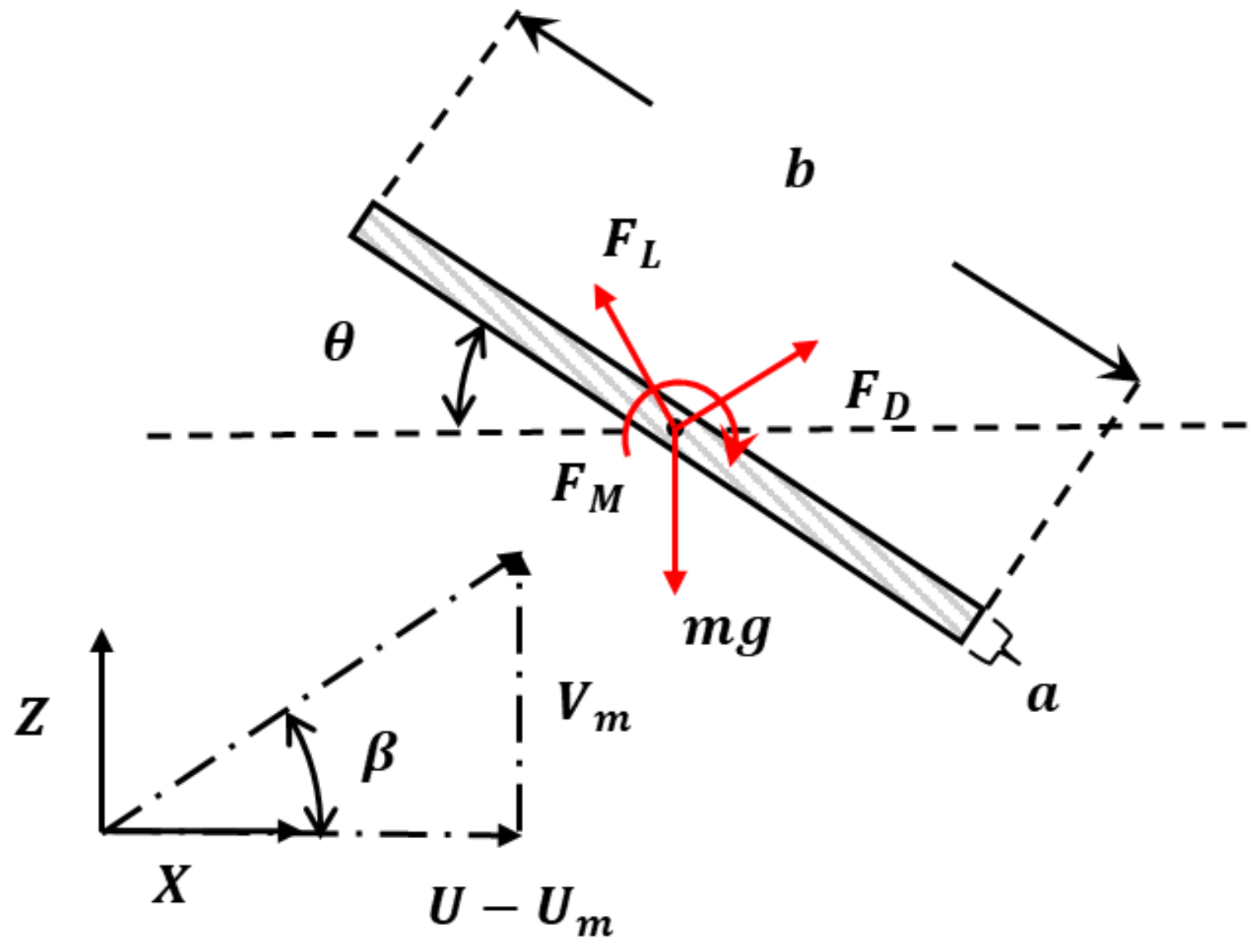

**Figure 4.** Aerodynamic-force, moment, and gravity components for a plate in the global Cartesian frame.

$$\frac{d^2x}{dt^2} = \frac{F_D \cos\beta - F_L \sin\beta}{m} = \frac{\rho_a (C_D \cos\beta - C_L \sin\beta)\left[(U-U_m)^2 + {V_m}^2\right]}{2a\rho_m} \quad \text{(Eq.4)}$$

$$\frac{d^2z}{dt^2} = \frac{F_D \sin\beta + F_L \cos\beta}{m} = \frac{\rho_a (C_D \sin\beta + C_L \cos\beta)\left[(U-U_m)^2 + {V_m}^2\right]}{2a\rho_m} - g \quad \text{(Eq.5)}$$

$$\frac{d^2\theta}{dt^2} = \frac{F_M}{I} = \frac{\rho_a}{2I} A_c L C_M\left[(U - U_m)^2 + {V_m}^2\right] \quad \text{(Eq.6)}$$

Tachikawa also developed a non-dimensional format for these equations as Eq. (8)-(10). In this dimensionless format, the physical attributes of the debris and the intensity of the incoming flow are represented by the Tachikawa parameter $K$ shown in Eq. (7). Wind tunnel testing can provide the estimation value of $K$ with different debris in various aspect ratio (J.D. Holmes et al., 2006;

Lin et al., 2007). However, the Tachikawa method only introduces one-dimensional flow and restricts the debris motion with three degrees of freedom, which is suitable for the wind tunnel but not appropriate in practical cases. Further, a three-dimensional model with overall consideration of lift, drag, side force, pitch moment, rolling moment, and deflection torque coefficients was established (Redmann et al., 1978). Thus, the following section discusses an alternative solution that is more suitable for practical applications.

$$K = \frac{\rho_a U^2 A_c}{2mg} \quad \textit{(Eq.7)}$$

$$\frac{d^2x}{dt^2} = Kg(C_D \cos\beta - C_L \sin\beta)\left[\left(1 - \frac{U_m}{U}\right)^2 + \left(\frac{V_m}{U}\right)^2\right] \quad \textit{(Eq.8)}$$

$$\frac{d^2z}{dt^2} = Kg(C_D \sin\beta + C_L \cos\beta)\left[\left(1 - \frac{U_m}{U}\right)^2 + \left(\frac{V_m}{U}\right)^2\right] - g \quad \textit{(Eq.9)}$$

$$\frac{d^2\theta}{dt^2} = K\frac{mgA_cL}{I}C_M\left[\left(1 - \frac{U_m}{U}\right)^2 + \left(\frac{V_m}{U}\right)^2\right] \quad \textit{(Eq.10)}$$

2.2 *Three-dimensional six-degree-of-freedom (6-DOF) general debris EOM under vortex flow*

In practical cases, wind-borne debris modeling should be done in 6-DOF with three degrees of translational motion freedom and three degrees of rotational freedom with angular speed. Recent research has established the 6-DOF general debris EOM under three-dimensional flow conditions (Abdelhady et al., 2021; Richards et al., 2008). In this 6-DOF EOM, the general debris is applied under the numerical wind field with resolved vortex flow components $(U, V, W)$, which will be discussed in Section 4, in the vortex cylindrical coordinate system $(R, \theta, Z)$.

2.2.1 *Coordinates transformation of the vortex flow field*

For computational convenience, the resolved vortex flow $(U, V, W)$ is calculated by the numerical vortex model with the debris position $(R_{m,i}, \theta_{m,i}, Z_{m,i},)$ under vortex cylindrical coordinates in Section 4. So, it is required to convert the vortex cylindrical coordinate system ($R, \theta, Z$) into the global reference Cartesian coordinates ($X, Y, Z$). At the time step $i$ under global coordinates, the coordinate of the vortex center is assumed as ($X_i, Y_i, Z_i$) and the mass center of debris is located at ($X_{m,i}, Y_{m,i}, Z_{m,i}$). Then, the relationship between the debris position under the vortex cylindrical coordinate ($R_{m,i}, \theta_{m,i}, Z_{m,i},$) and global coordinates ($X_{m,i}, Y_{m,i}, Z_{m,i}$) can be denoted as Eq. (11). Further, Then, the resolved vortex flow under the global coordinates ($V_X, V_Y, V_Z$) can be transformed as Eq. (12). As shown in Figure 5, the relative velocity vector for the debris under global coordinates are $(V_X - \dot{X}_{m,i})$, $(V_Y - \dot{Y}_{m,i})$, and $(V_Z - \dot{Z}_{m,i})$, which will be used for calculating the external force for the translational motion of the debris in the following.

$$\begin{bmatrix} X_{m,i} \\ Y_{m,i} \\ Z_{m,i} \end{bmatrix} = \begin{bmatrix} R_{m,i} \cos \theta_{m,i} \\ R_{m,i} \sin \theta_{m,i} \\ z_{m,i} \end{bmatrix} + \begin{bmatrix} X_i \\ Y_i \\ 0 \end{bmatrix} \quad \text{(Eq.11)}$$

$$\begin{bmatrix} V_X \\ V_Y \\ V_Z \end{bmatrix} = \begin{bmatrix} \cos\theta & -\sin\theta & 0 \\ \sin\theta & \cos\theta & 0 \\ 0 & 0 & 1 \end{bmatrix} \begin{bmatrix} U \\ V \\ W \end{bmatrix} \quad \text{(Eq.12)}$$

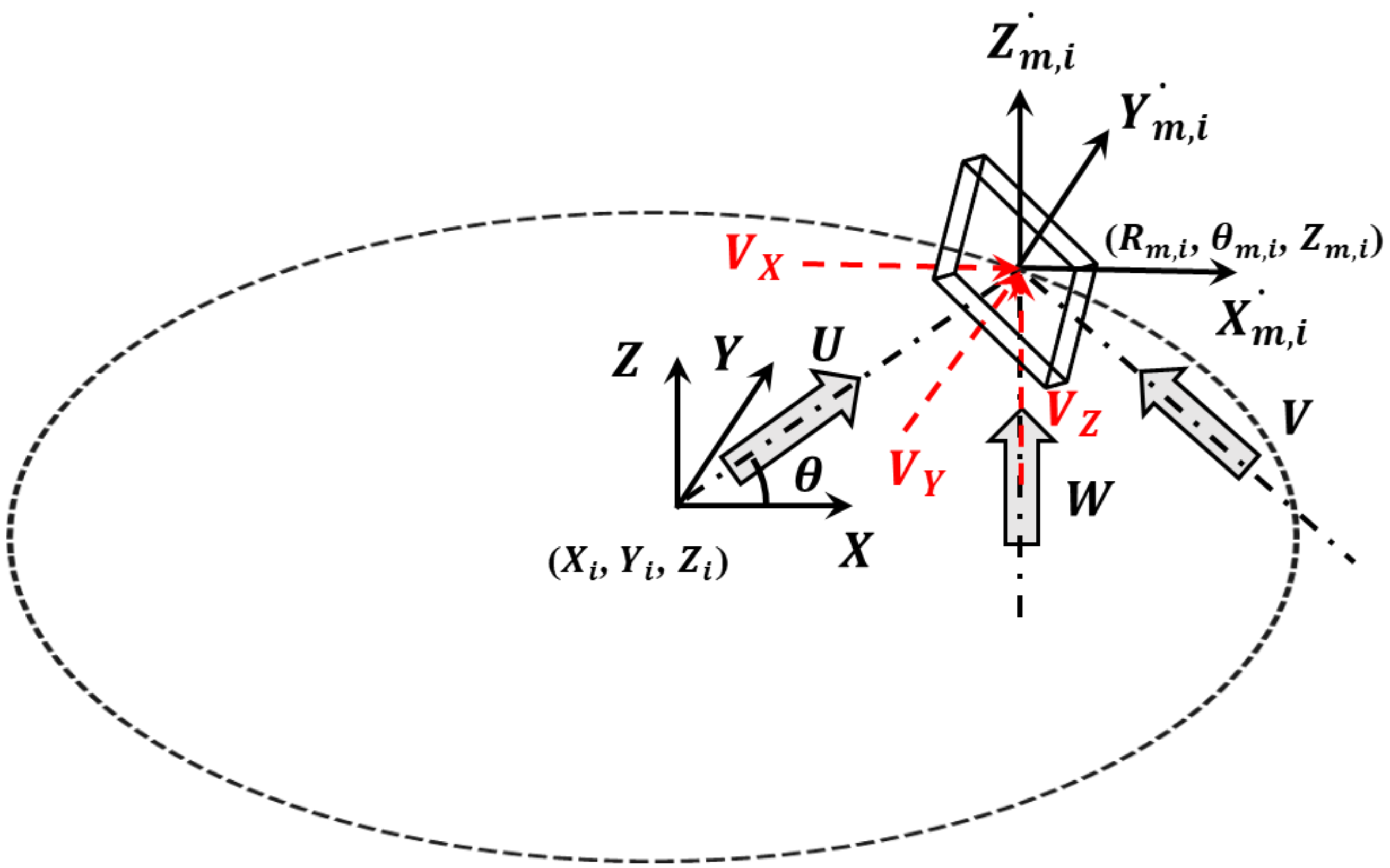

**Figure 5.** The schematic drawing of resolved wind flow components ($U$, $V$, $W$) and the transformed flow components ($V_X$, $V_Y$, $V_Z$) in the global coordinates. The debris translational motion speeds are denoted in time derivatives as ($\dot{X}_{m,i}$, $\dot{Y}_{m,i}$, $\dot{Z}_{m,i}$) in the global coordinates.

### 2.2.2 *Debris rotation under rigid body coordinates*

A rigid body coordinates with the orthogonal axes ($X^R$, $Y^R$, $Z^R$) along the principal axes of inertia ($I_{XX}$, $I_{YY}$, $I_{ZZ}$) of the debris is required to describe the rotation of the debris. Figure 6 shows the rotation from the global Cartesian coordinates in Case (a) to the rigid body coordinates in Case (b) by introducing the Euler angles ($\theta_{roll}$, $\theta_{pitch}$, $\theta_{yaw}$). Also, for posture control, the debris in the rigid body coordinates is fixed as the thickness length. $a$ along the principal axis $I_{XX}$, the vertical length $b$ along the principal axis $I_{ZZ}$, and the horizontal length $c$ along the principal axis $I_{YY}$. To specify the orientation of the debris, $a < b < c$ should be satisfied with the lengths of the debris.

Based on the definition of the Euler angle, the coordinates rotation process should be developed in the following sequence. Firstly, in the global Cartesian coordinates, the $X$-$Y$plane is rotated around $Z$-axis with the yaw angle $\theta_{yaw}$. Secondly, the whole coordinates rotate around the new. $Y$-axis with the pitch angle $\theta_{pitch}$. Then, the whole coordinates rotate around the new $X$-axis, which is also the $X^R$-axis in the rigid body coordinates, with the roll angle $\theta_{roll}$. As a result, the reference coordinates have been rotated from Figure 6. (a) to Figure 6. (b), and the rotation process can be denoted in Eq. (13) with a rotational matrix $\boldsymbol{R_x}(\theta_{roll})$, $\boldsymbol{R_y}(\theta_{pitch})$, $\boldsymbol{R_z}(\theta_{yaw})$ for the coordinates transform of the Euler angles ($\theta_{roll}$, $\theta_{pitch}$, $\theta_{yaw}$), respectively.

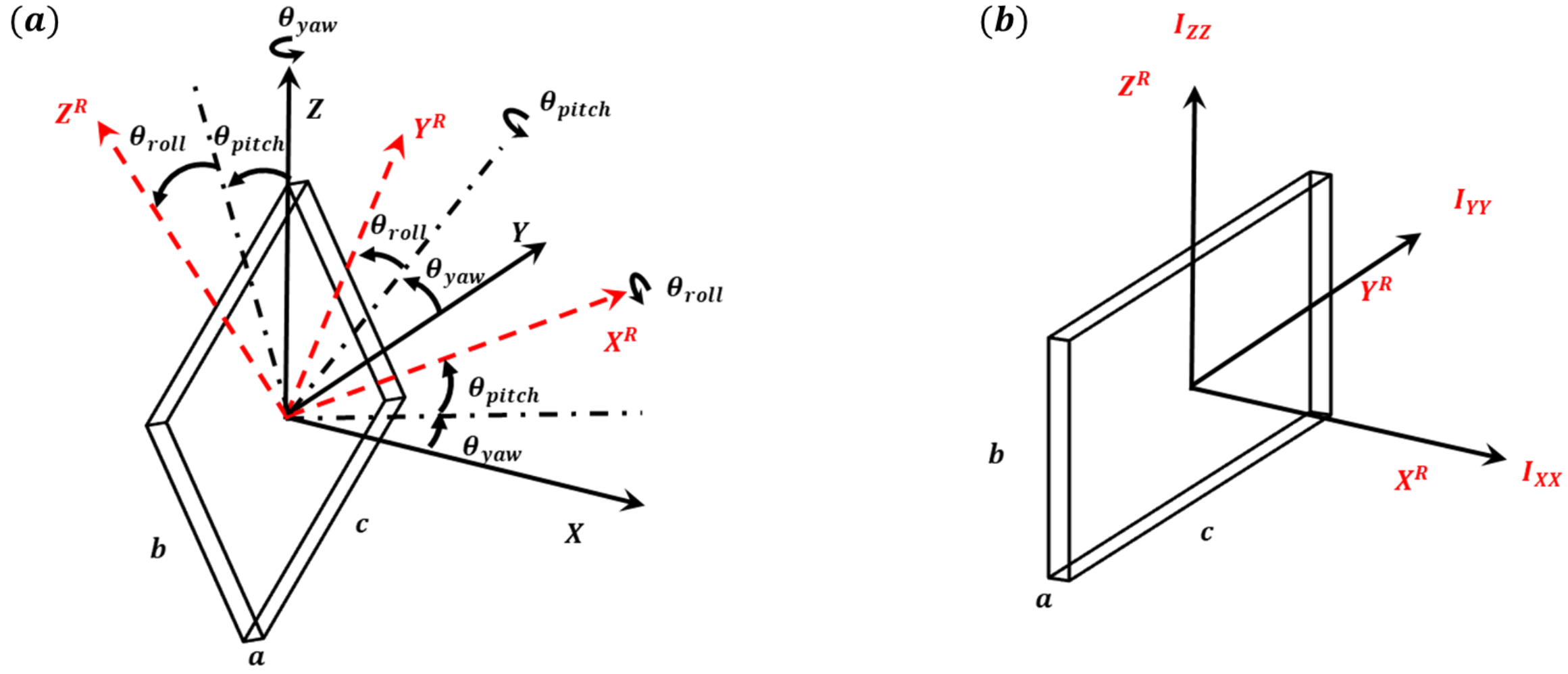

**Figure 6.** (a) The rotated debris rigid body coordinates with the Euler angle ($\theta_{yaw}$, $\theta_{pitch}$, $\theta_{roll}$) in the global Cartesian coordinates. (b) The debris rigid body coordinates with three principal axes of inertia ($I_{XX}$, $I_{YY}$, $I_{ZZ}$) as the orthogonal axes ($X^R$, $Y^R$, $Z^R$)

$$\begin{bmatrix} X^R \\ Y^R \\ Z^R \end{bmatrix} = \boldsymbol{R_z}(\theta_{yaw})\boldsymbol{R_y}(\theta_{pitch})\boldsymbol{R_x}(\theta_{roll}) \begin{bmatrix} X \\ Y \\ Z \end{bmatrix} =$$

$$\begin{bmatrix} cos(\theta_{yaw}) & -sin(\theta_{yaw}) & 0 \\ sin(\theta_{yaw}) & cos(\theta_{yaw}) & 0 \\ 0 & 0 & 1 \end{bmatrix} \begin{bmatrix} cos(\theta_{pitch}) & 0 & sin(\theta_{pitch}) \\ 0 & 1 & 0 \\ -sin(\theta_{pitch}) & 0 & cos(\theta_{pitch}) \end{bmatrix} \begin{bmatrix} 1 & 0 & 0 \\ 0 & cos(\theta_{roll}) & -sin(\theta_{roll}) \\ 0 & sin(\theta_{roll}) & cos(\theta_{roll}) \end{bmatrix} \begin{bmatrix} X \\ Y \\ Z \end{bmatrix}$$

*(Eq.13)*

Similarly, the rotation process from the rigid body coordinates to the global Cartesian coordinates can be calculated by reversing the rotation process discussed above. Then, the transformation matrix is denoted as Eq. (14).

$$\begin{bmatrix} X \\ Y \\ Z \end{bmatrix} = \boldsymbol{R_x}(\theta_{roll})^T \boldsymbol{R_y}(\theta_{pitch})^T \boldsymbol{R_z}(\theta_{yaw})^T \begin{bmatrix} X^R \\ Y^R \\ Z^R \end{bmatrix} =$$

$$\begin{bmatrix} 1 & 0 & 0 \\ 0 & cos(\theta_{roll}) & sin(\theta_{roll}) \\ 0 & -sin(\theta_{roll}) & cos(\theta_{roll}) \end{bmatrix} \begin{bmatrix} cos(\theta_{pitch}) & 0 & -sin(\theta_{pitch}) \\ 0 & 1 & 0 \\ sin(\theta_{pitch}) & 0 & cos(\theta_{pitch}) \end{bmatrix} \begin{bmatrix} cos(\theta_{yaw}) & sin(\theta_{yaw}) & 0 \\ -sin(\theta_{yaw}) & cos(\theta_{yaw}) & 0 \\ 0 & 0 & 1 \end{bmatrix} \begin{bmatrix} X^R \\ Y^R \\ Z^R \end{bmatrix}$$

*(Eq.14)*

Thus, the 6-DOF EOM of the debris can be developed in the rigid body coordinates and the relative coordinates can be converted from the global coordinates with the given Euler angles ($\theta_{roll}$, $\theta_{pitch}$, $\theta_{yaw}$). In addition, the Euler angles also represent the debris posture and provide a convenient format for calculating the aerodynamic effects along the principal axes as following section.

### 2.2.3 *6-DOF equations of motion in the debris body frame*

Once the conversion from the vortex polar coordinate systems ($r$, $\theta$, $z$) to the global reference coordinates ($X$, $Y$, $Z$) and the conversion from the global reference coordinates ($X$, $Y$, $Z$) to the debris rigid body coordinates ($X^R$, $Y^R$, $Z^R$) have been established, the 6-DOF EOM for the debris can be formulated in its rigid-body coordinate system. In Figure 7, the relative velocity is combined with the orthogonal speed vectors $(V_X - \dot{X}_{m,\iota})$, $(V_Y - \dot{Y}_{m,\iota})$, and $(V_Z - \dot{Z}_{m,\iota})$ with the projection angle $\gamma$ and tilt angle $\varepsilon$ denoted in the global Cartesian coordinates as Eq. (15). Then, in the force analysis diagram, the normal aerodynamic forces of $F_{XX}$, $F_{YY}$, $F_{ZZ}$, and gravity is considered to generate the acceleration of the debris's translational motion. Further, the moment $M_{XX}$, $M_{YY}$, $M_{ZZ}$ around the principal will provide the angular acceleration. Then, based on Eq. (15), the relative flow components ($V_{XX}$, $V_{YY}$, $V_{ZZ}$) under the rigid body frame can be transformed from relative wind speeds in the global coordinates as Eq. (16). Further, the quasi-steady normal forces can be expressed along the principal axes with corresponding cross-section area and normal force coefficients ($C_{FXX}$, $C_{FYY}$, $C_{FZZ}$) as Eq. (17). The estimation of the normal force coefficients will be discussed in Section 3. In addition, gravity also contributes to the acceleration of the debris, then, the gravity is also converted into the resolved vector under the rigid body coordinates as Eq. (18). Then, the translational EOM of the debris can be denoted in Eq. (19) under the rigid body coordinates.

$$\begin{bmatrix} \gamma \\ \varepsilon \end{bmatrix} = tan^{-1} \begin{bmatrix} \frac{V_Z - \dot{Z}_{m,i}}{\sqrt{(V_X - \dot{X}_{m,i})^2 + (V_Y - \dot{Y}_{m,i})^2}} \\ \frac{V_Y - \dot{Y}_{m,i}}{V_X - \dot{X}_{m,i}} \end{bmatrix} \tag{Eq.15}$$

$$\begin{bmatrix} V_{XX} \\ V_{YY} \\ V_{ZZ} \end{bmatrix} = \boldsymbol{R_z}(\theta_{yaw})\boldsymbol{R_y}(\theta_{pitch})\boldsymbol{R_x}(\theta_{roll}) \begin{bmatrix} V_X - \dot{X}_{m,i} \\ V_Y - \dot{Y}_{m,i} \\ V_Z - \dot{Z}_{m,i} \end{bmatrix} \tag{Eq.16}$$

$$\begin{bmatrix} F_{XX} \\ F_{YY} \\ F_{ZZ} \end{bmatrix} = \frac{\rho_a}{2}\left({V_{XX}}^2 + {V_{YY}}^2 + {V_{ZZ}}^2\right) \begin{bmatrix} bcC_{FXX} \\ abC_{FYY} \\ acC_{FZZ} \end{bmatrix} \tag{Eq.17}$$

$$\begin{bmatrix} G_{XX} \\ G_{YY} \\ G_{ZZ} \end{bmatrix} = \boldsymbol{R_z}(\theta_{yaw})\boldsymbol{R_y}(\theta_{pitch})\boldsymbol{R_x}(\theta_{roll}) \begin{bmatrix} 0 \\ 0 \\ -mg \end{bmatrix} \tag{Eq.18}$$

$$\begin{bmatrix} \ddot{X}_m^R \\ \ddot{Y}_m^R \\ \ddot{Z}_m^R \end{bmatrix} = \frac{1}{m} \begin{bmatrix} F_{XX} + G_{XX} \\ F_{YY} + G_{YY} \\ F_{ZZ} + G_{ZZ} \end{bmatrix} \tag{Eq.19}$$

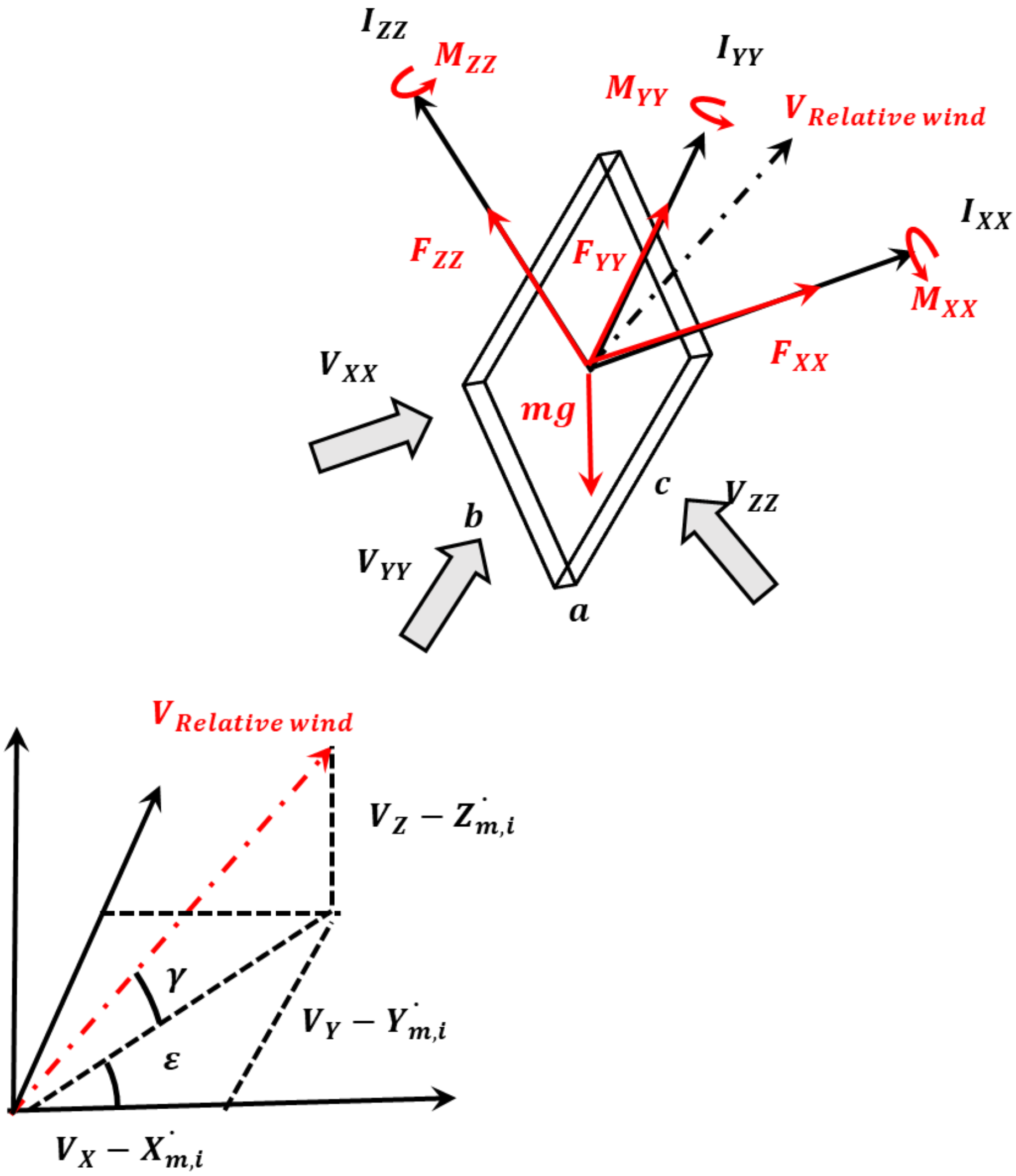

**Figure 7.** Relative-flow components, aerodynamic forces and moments, and gravity resolved in the debris body-fixed frame.

Based on Euler's equation shown in Eq. (20), the external torque of the rigid body fixed frame $\boldsymbol{M}$ can be expressed with the momentum of inertia of the principal axes $\boldsymbol{I}$, angular velocity $\boldsymbol{\omega}$ and the time derivatives of angular velocity $\dot{\boldsymbol{\omega}}$. Analogously, as shown in Eq. (21), the external torque in the rigid body fixed frame ($M_{XX}$, $M_{YY}$, and $M_{ZZ}$) are defined with the moment coefficients ($C_{MXX}$, $C_{MYY}$, $C_{MZZ}$) under the maximum cross-section area (the smallest section area $ab$ is neglected) with the largest length as the torque radius, and the estimation of the coefficients will be introduced in section 3. Then, the time derivatives of angular velocity $\dot{\boldsymbol{\omega}}$ around each principal axis can be computed as the ordinary differential equations (ODEs) in Eq. (22), and the numerical solution of $\dot{\boldsymbol{\omega}}$ and $\boldsymbol{\omega}$ are developed in Section 5. Further, the angular velocity vector $\boldsymbol{\omega}$ can be expressed in terms of the Euler Angles in Eq. (23). In this equation, ($\dot{\boldsymbol{\theta}}_{\boldsymbol{roll}}$, $\dot{\boldsymbol{\theta}}_{\boldsymbol{pitch}}$, $\dot{\boldsymbol{\theta}}_{\boldsymbol{yaw}}$) are the time derivatives of the Euler angles, ($\widehat{\boldsymbol{I}_{\boldsymbol{XX}}}$, $\widehat{\boldsymbol{I}_{\boldsymbol{YY}}}$, $\widehat{\boldsymbol{I}_{\boldsymbol{ZZ}}}$) are the unit vectors, respectively. Correspondingly, the Euler angular velocity ($\dot{\boldsymbol{\theta}}_{\boldsymbol{roll}}$, $\dot{\boldsymbol{\theta}}_{\boldsymbol{pitch}}$, $\dot{\boldsymbol{\theta}}_{\boldsymbol{yaw}}$), as shown in Eq. (24), can be computed by inverting the kinematic matrix used in Eq. (23). Also, the numerical solution of the Euler angles ($\theta_{roll}$, $\theta_{pitch}$, $\theta_{yaw}$) and the corresponding time derivatives are developed in Section 5.

$$\boldsymbol{M} = \boldsymbol{I}\dot{\boldsymbol{\omega}} + \boldsymbol{\omega} \times (\boldsymbol{I}\boldsymbol{\omega}) = \begin{bmatrix} I_{XX}\,\dot{\omega}_{XX} \\ I_{YY}\,\dot{\omega}_{YY} \\ I_{ZZ}\,\dot{\omega}_{ZZ} \end{bmatrix} + \begin{bmatrix} (I_{ZZ} - I_{YY})\omega_{YY}\omega_{ZZ} \\ (I_{XX} - I_{ZZ})\omega_{ZZ}\omega_{XX} \\ (I_{YY} - I_{XX})\omega_{XX}\omega_{YY} \end{bmatrix}$$

$$\textit{Where } \boldsymbol{M} = \begin{bmatrix} M_{XX} \\ M_{YY} \\ M_{ZZ} \end{bmatrix}, \boldsymbol{I} = \begin{bmatrix} I_{XX} & 0 & 0 \\ 0 & I_{YY} & 0 \\ 0 & 0 & I_{ZZ} \end{bmatrix}, \boldsymbol{\omega} = \begin{bmatrix} \omega_{XX} \\ \omega_{YY} \\ \omega_{ZZ} \end{bmatrix} \qquad \textit{(Eq.20)}$$

$$\begin{bmatrix} M_{XX} \\ M_{YY} \\ M_{ZZ} \end{bmatrix} = \frac{\rho_a}{2}\left({V_{XX}}^2 + {V_{YY}}^2 + {V_{ZZ}}^2\right)\begin{bmatrix} ac^2 C_{MXX} \\ cb^2 C_{MYY} \\ bc^2 C_{MZZ} \end{bmatrix}$$

*(Eq.21)*

$$\begin{bmatrix} \dot{\omega}_{XX} \\ \dot{\omega}_{YY} \\ \dot{\omega}_{ZZ} \end{bmatrix} = \begin{bmatrix} \frac{M_{XX}}{I_{XX}} \\ \frac{M_{YY}}{I_{YY}} \\ \frac{M_{ZZ}}{I_{ZZ}} \end{bmatrix} - \begin{bmatrix} \left(\frac{I_{ZZ}-I_{YY}}{I_{XX}}\right)\omega_{YY}\omega_{ZZ} \\ \left(\frac{I_{XX}-I_{ZZ}}{I_{YY}}\right)\omega_{ZZ}\omega_{XX} \\ \left(\frac{I_{YY}-I_{XX}}{I_{ZZ}}\right)\omega_{XX}\omega_{YY} \end{bmatrix} \qquad \textit{(Eq.22)}$$

$$\begin{bmatrix} \omega_{XX} \\ \omega_{YY} \\ \omega_{ZZ} \end{bmatrix} = \dot{\boldsymbol{\theta}}_{\boldsymbol{roll}}\widehat{\boldsymbol{I}_{\boldsymbol{XX}}} + \dot{\boldsymbol{\theta}}_{\boldsymbol{pitch}}\widehat{\boldsymbol{I}_{\boldsymbol{YY}}} + \dot{\boldsymbol{\theta}}_{\boldsymbol{yaw}}\widehat{\boldsymbol{I}_{\boldsymbol{ZZ}}} = \begin{bmatrix} 1 & 0 & -sin(\theta_{pitch}) \\ 0 & cos(\theta_{roll}) & cos(\theta_{pitch})\,sin(\theta_{roll}) \\ 0 & -sin(\theta_{roll}) & cos(\theta_{pitch})\,cos(\theta_{roll}) \end{bmatrix}\begin{bmatrix} \dot{\theta}_{roll} \\ \dot{\theta}_{pitch} \\ \dot{\theta}_{yaw} \end{bmatrix}$$

*(Eq.23)*

$$\begin{bmatrix} \dot{\theta_{roll}} \\ \dot{\theta_{pitch}} \\ \dot{\theta_{yaw}} \end{bmatrix} = \begin{bmatrix} 1 & 0 & -\sin(\theta_{pitch}) \\ 0 & \cos(\theta_{roll}) & \cos(\theta_{pitch})\sin(\theta_{roll}) \\ 0 & -\sin(\theta_{roll}) & \cos(\theta_{pitch})\cos(\theta_{roll}) \end{bmatrix}^{-1} \begin{bmatrix} \omega_{XX} \\ \omega_{YY} \\ \omega_{ZZ} \end{bmatrix} =$$

$$\begin{bmatrix} \omega_{XX} + \tan(\theta_{pitch})\,(\omega_{YY}\sin(\theta_{roll}) + \omega_{ZZ}\cos(\theta_{roll})) \\ \omega_{YY}\cos(\theta_{roll}) - \omega_{ZZ}\sin(\theta_{roll}) \\ \frac{\omega_{ZZ}\cos(\theta_{roll}) + \omega_{YY}\sin(\theta_{roll})}{\cos(\theta_{pitch})} \end{bmatrix} \quad (Eq.24)$$

Considering debris types, the plate debris should be employed with 6-DOF EOM, but the other debris should have a reduced number of the DOF of the EOM. For example, in Figure 8. (a), a compact object may be represented by three translational coordinates if orientation-dependent forces are neglected. Then, the 6-DOF EOM can be reduced to the 3-DOF EOM $[\ddot{X_m} \quad \ddot{Y_m} \quad \ddot{Z_m}]^T$. As for the rod debris in Figure 8. (b), For a slender rod, rotation about its long principal axis $I_{YY}$ may be omitted, giving a five-state representation. Then, the 5-DOF EOM should be $[\ddot{X_m} \quad \ddot{Y_m} \quad \ddot{Z_m} \quad \dot{\omega_{XX}} \quad \dot{\omega_{ZZ}}]^T$ in the rigid body coordinates.

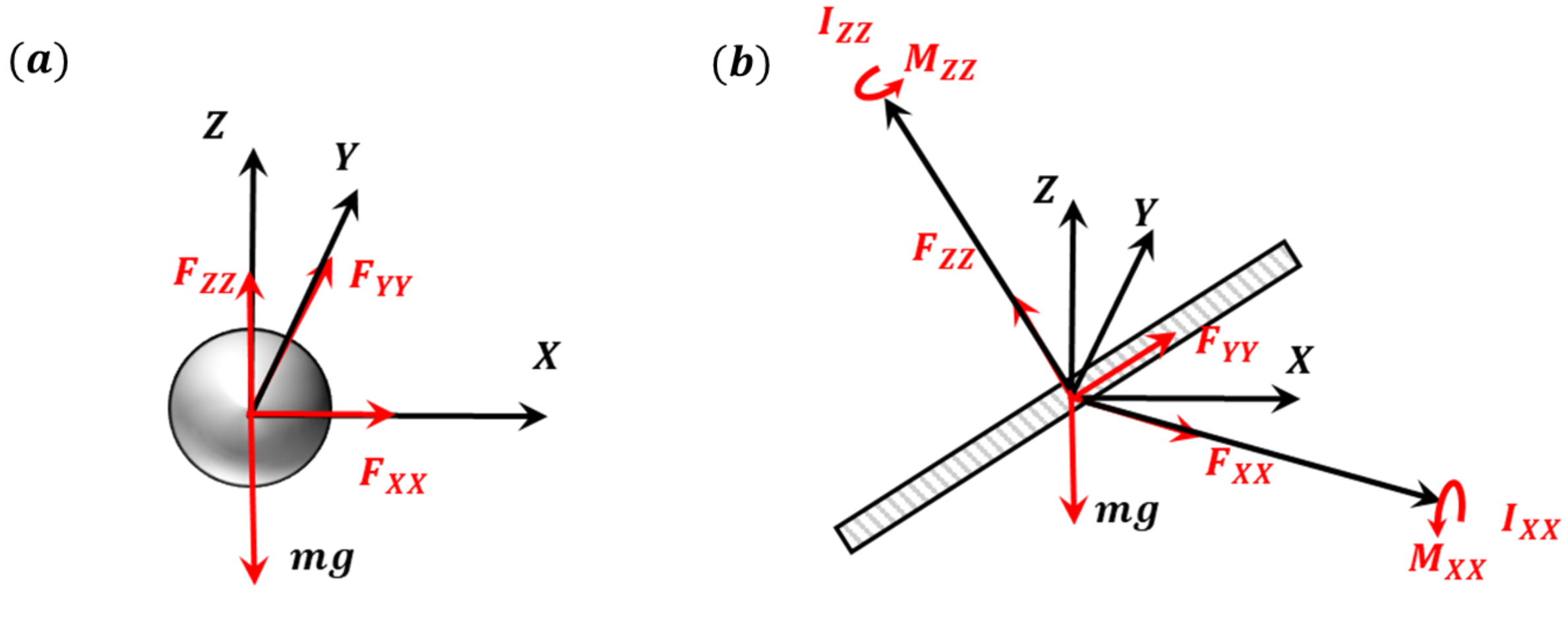

**Figure 8.** Reduced representations for (a) compact and (b) rod-like debris. The plate application retains six degrees of freedom.

# 3 DEBRIS AERODYNAMIC FORCES AND COEFFICIENTS

To obtain the aerodynamic coefficients for the windborne debris, wind tunnel tests and Computational Fluid Dynamics (CFD) simulation are ideal options to provide an estimation. As for wind tunnel tests, (Richards et al., 2008) tested rectangular plates at various wind attack angles ($\gamma$) and tilt angles ($\varepsilon$). Also, CFD has already been applied in recent studies of simulating the windborne debris trajectories, and unsteady/quasi-steady flow methods are the two main simulation methods applied in the computational fluid-dynamics rigid body (CFD-RBD) method to solve the numerical trajectories (Kakimpa et al., 2012a, 2012b; Saini and Shafei, 2022). In the unsteady flow simulation method, the debris motion in the wind field is considered a Fluid-Structure Interaction (FSI) problem, and Large Eddy Simulation (LES) with dynamic mesh technique is applied for solving the time-varying debris spatial position (Xu et al., 2020).In the steady flow simulation, Kakimpa et al., (2012) stated that the debris trajectories derived from CFD-RBD simulations align with those predicted by quasi-steady analytical methods after the initial stage of flight.

The aerodynamic approximation resolves a body with dimensions $(a, b, c)$ into the $X$-$Y$, $X$-$Z$, and $Y$-$Z$ planes. Each steady CFD domain extends approximately ten characteristic lengths in the streamwise direction and four in each transverse direction, summarized as $(10c, 4c, 4c)$. These dimensions place external boundaries away from the plate while keeping the coefficient sweep tractable; no retained domain-independence study was available for this revision. Each plane returns drag, lift, and moment coefficients. For the $X$-$Y$ plane, the relative velocity $V_{XY}$ is assembled from $V_{XX}$ and $V_{YY}$, with attack angle $\gamma_{XY}$. Forces and moment are transformed to body-axis quantities. Eq. (25) combines the three planar solutions and subtracts duplicated axial contributions. Eq. (26) expresses the forces through projected areas, and Eq. (27) gives the body-axis force and moment coefficients. The earlier term momentum coefficient has been corrected to moment coefficient.

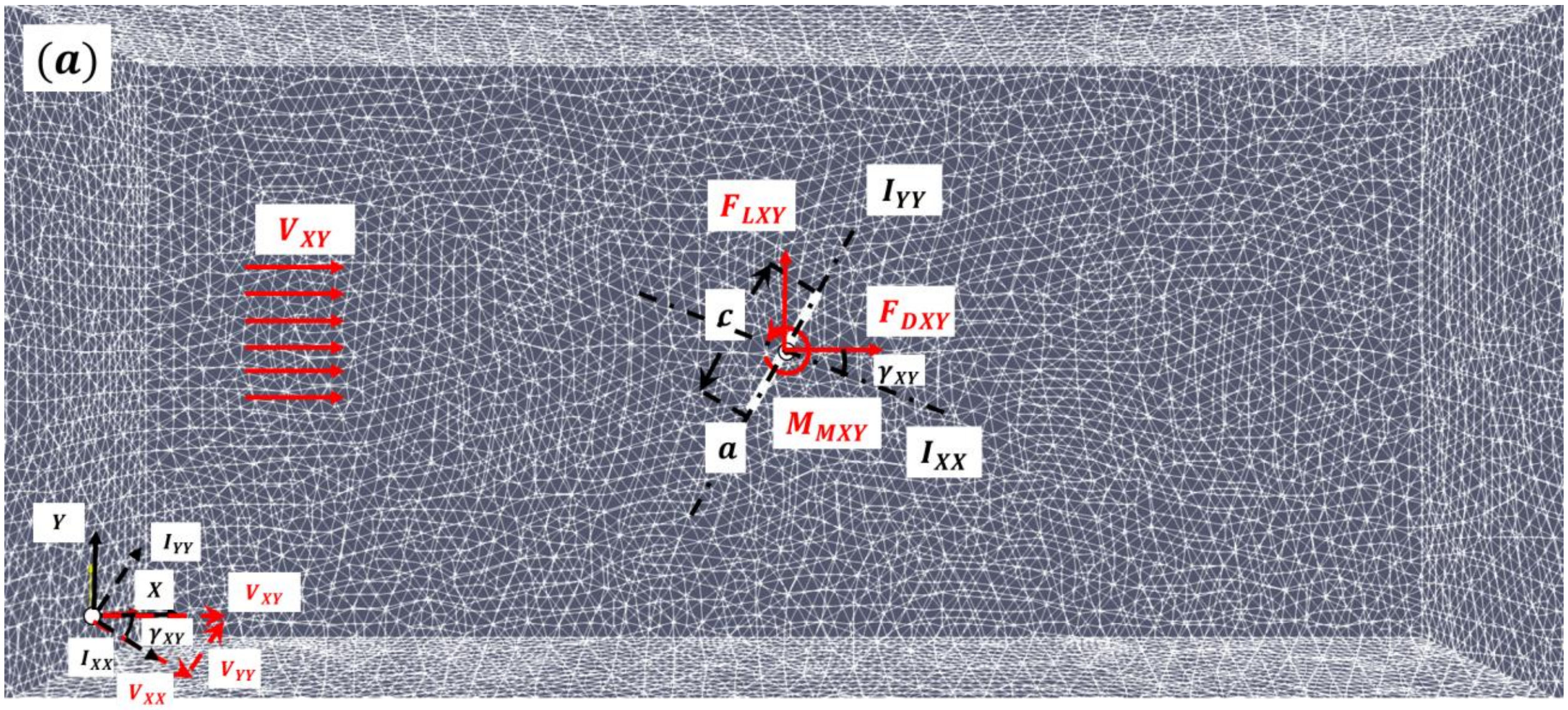

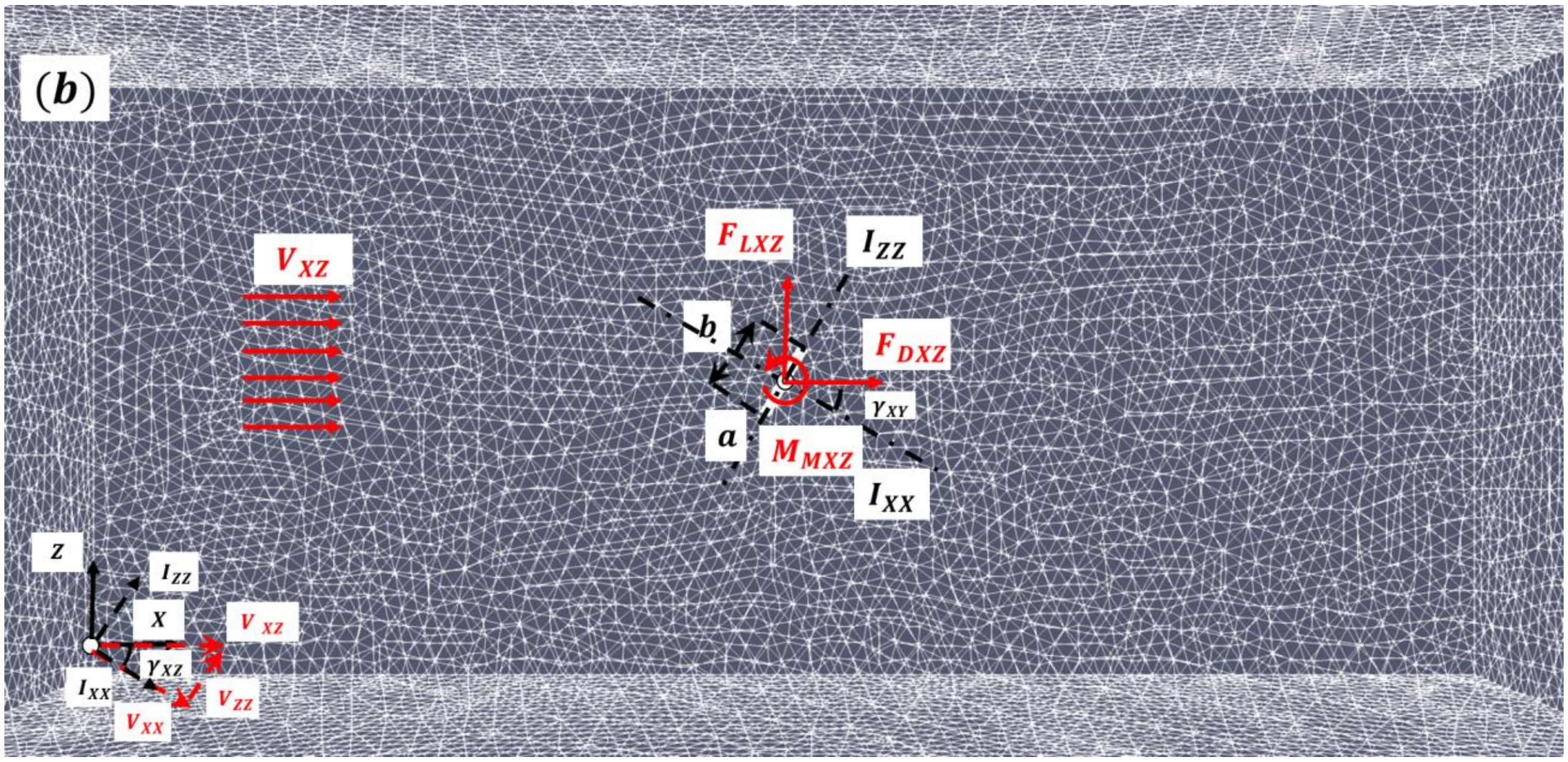

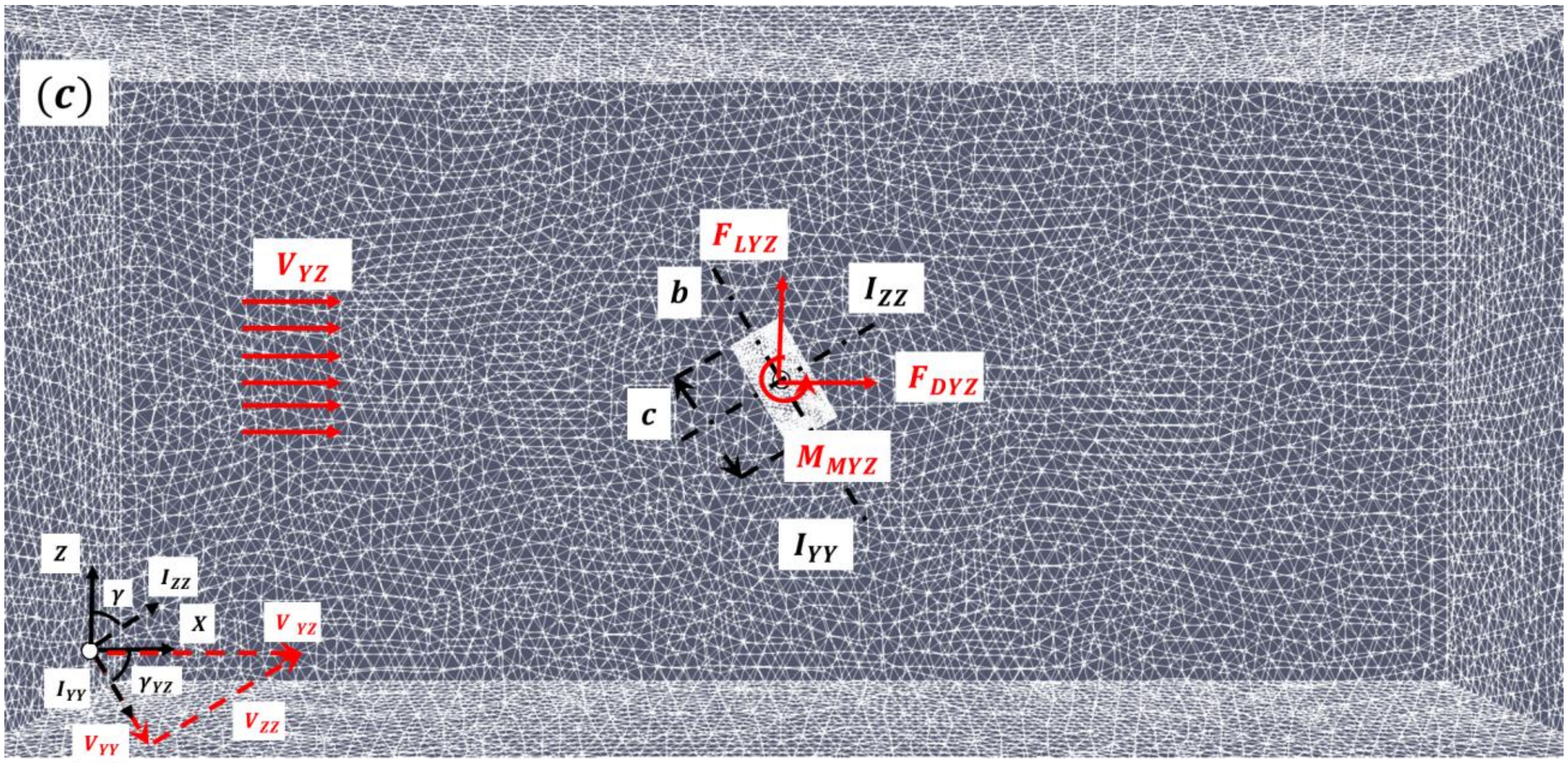

**Figure 9.** Planar CFD calculations of drag, lift, and moment in the (a) $X$-$Y$, (b) $X$-$Z$, and (c) $Y$-$Z$ planes.

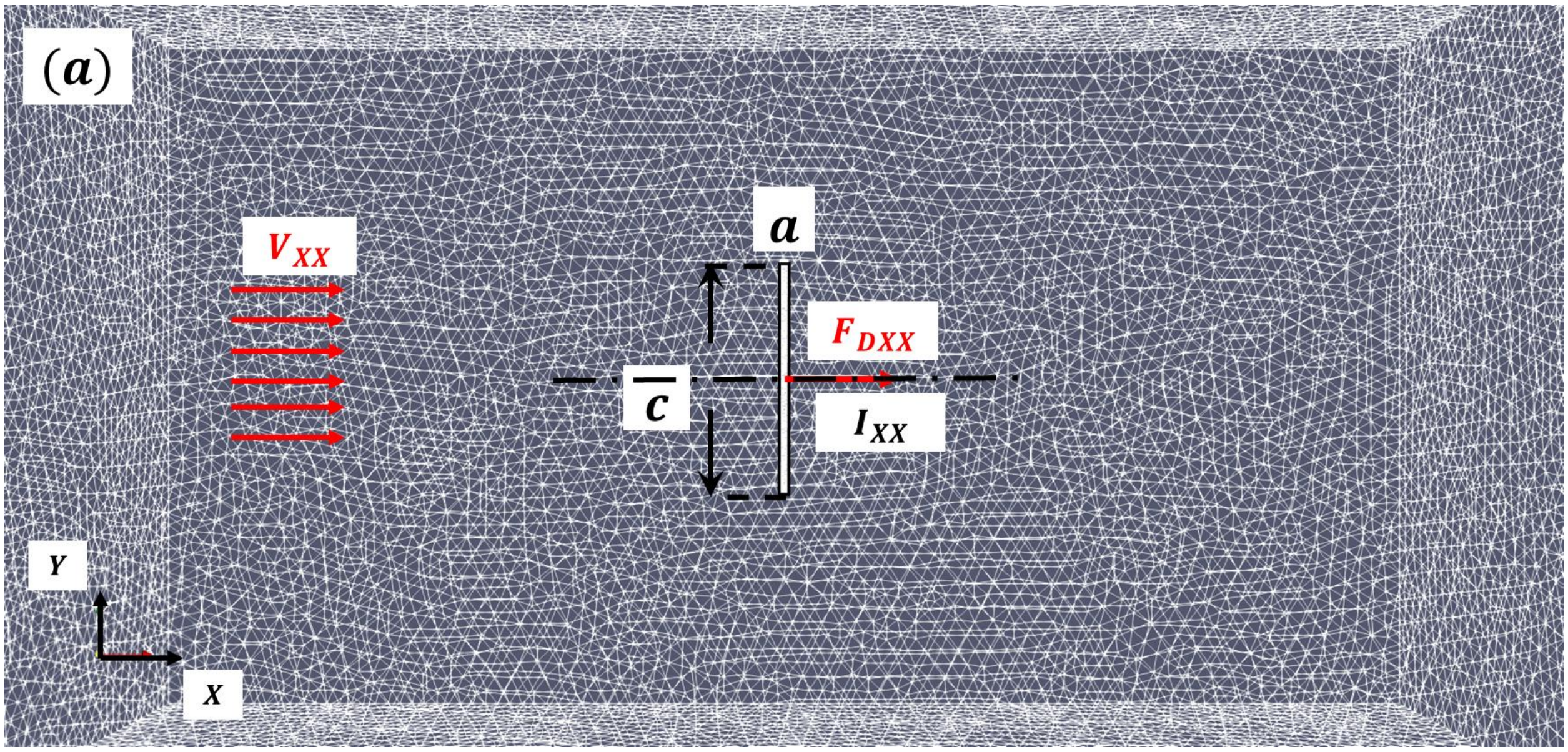

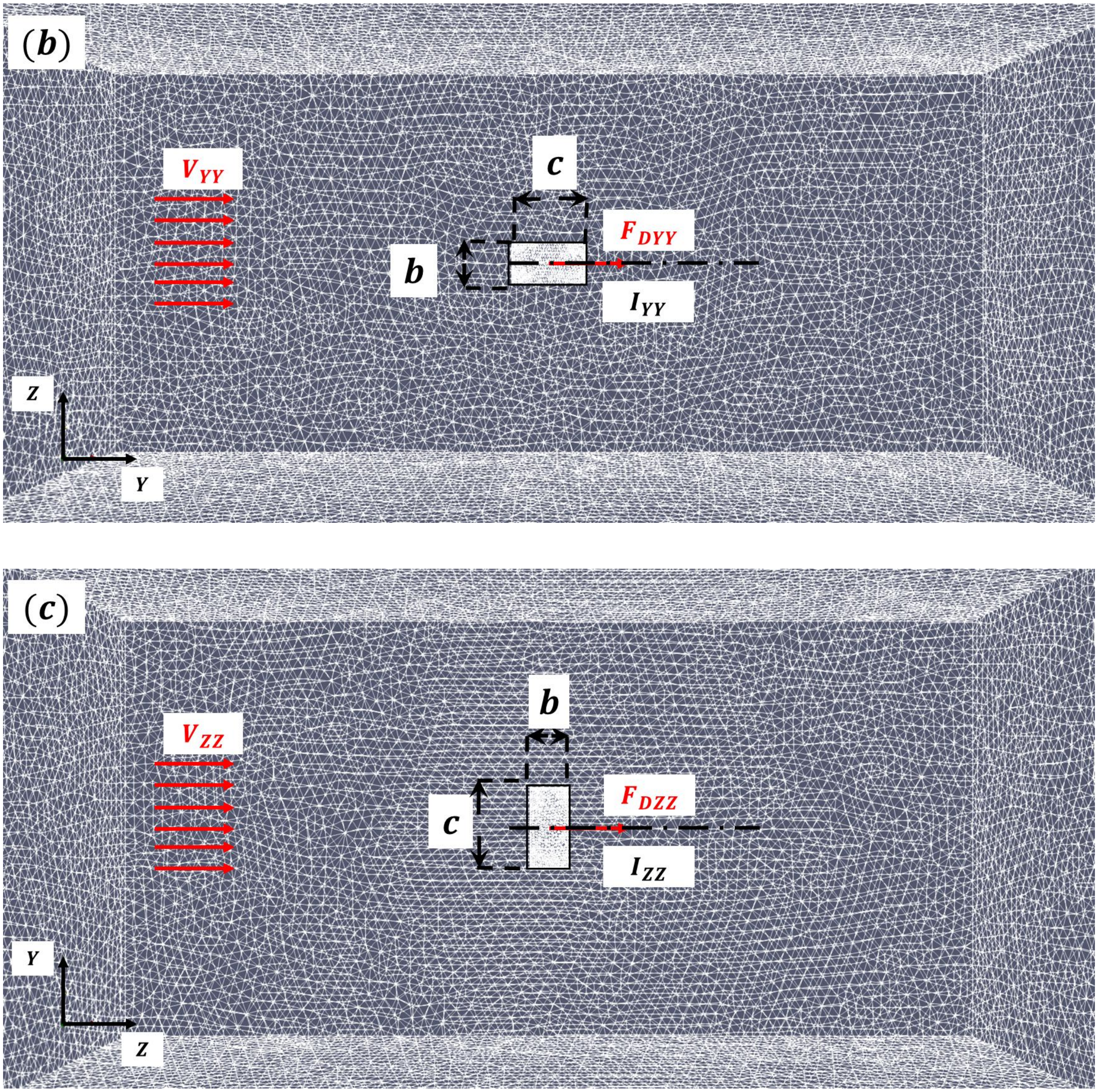

**Figure 10.** Axial compensation terms used when combining the (a) $X$-$Y$ and $X$-$Z$, (b) $X$-$Y$ and $Y$-$Z$, and (c) $X$-$Z$ and $Y$-$Z$ planar solutions .

$$F_{XX} = (F_{DXY}\cos\gamma_{XY} - F_{LXY}\sin\gamma_{XY}) + (F_{DXZ}\cos\gamma_{XZ} - F_{LXZ}\sin\gamma_{XZ}) - F_{DXX} \quad \textit{(Eq.25a)}$$

$$F_{YY} = (F_{DXY}\sin\gamma_{XY} + F_{LXY}\cos\gamma_{XY}) + (F_{DYZ}\cos\gamma_{YZ} - F_{LYZ}\sin\gamma_{YZ}) - F_{DYY} \quad \textit{(Eq.25b)}$$

$$F_{ZZ} = (F_{DXZ}\sin\gamma_{XZ} + F_{LXZ}\sin\gamma_{XZ}) + (F_{DYZ}\sin\gamma_{YZ} + F_{LYZ}\cos\gamma_{YZ}) - F_{DZZ} \quad \textit{(Eq.25c)}$$

$$M_{XX} = M_{MYZ}, \qquad M_{YY} = M_{MXZ}, \qquad M_{ZZ} = M_{MXY} \quad \textit{(Eq.25d)}$$

$$\mathbf{Q}_a = \frac{\rho_a}{2} V_R^2 [bcC_{FXX} \quad abC_{FYY} \quad acC_{FZZ} \quad ac^2 C_{MXX} \quad cb^2 C_{MYY} \quad bc^2 C_{MZZ}]^T,$$

$$V_R^2 = V_{XX}^2 + V_{YY}^2 + V_{ZZ}^2 \quad \textit{(Eq.26)}$$

$$\lambda_{XY} = \frac{V_{XX}^2 + V_{YY}^2}{V_R^2}, \quad \lambda_{XZ} = \frac{V_{XX}^2 + V_{ZZ}^2}{V_R^2}, \quad \lambda_{YZ} = \frac{V_{YY}^2 + V_{ZZ}^2}{V_R^2} \quad \textit{(Eq.27a)}$$

$$\lambda_{XX} = \frac{V_{XX}^2}{V_R^2}, \quad \lambda_{YY} = \frac{V_{YY}^2}{V_R^2}, \quad \lambda_{ZZ} = \frac{V_{ZZ}^2}{V_R^2} \quad \textit{(Eq.27b)}$$

$$C_{FXX} = \lambda_{XY}(C_{DXY}\cos\gamma_{XY} - C_{LXY}\sin\gamma_{XY}) + \lambda_{XZ}(C_{DXZ}\cos\gamma_{XZ} - C_{LXZ}\sin\gamma_{XZ}) - \lambda_{XX}C_{DXX}$$

*(Eq.27c)*

$$C_{FYY} = \lambda_{XY}\frac{c}{a}(C_{DXY}\sin\gamma_{XY} + C_{LXY}\cos\gamma_{XY}) + \lambda_{YZ}\frac{c}{b}(C_{DYZ}\cos\gamma_{XZ} - C_{LYZ}\sin\gamma_{YZ}) - \lambda_{YY}C_{DYY}$$

*(Eq.27d)*

$$C_{FZZ} = \lambda_{XZ}\frac{b}{a}(C_{DXY}\cos\gamma_{XY} - C_{LXY}\sin\gamma_{XY}) + \lambda_{YZ}(C_{DYZ}\sin\gamma_{XZ} + C_{LYZ}\cos\gamma_{XZ}) - \lambda_{ZZ}C_{DZZ}$$

*(Eq.27e)*

$$C_{MXX} = \lambda_{YZ}C_{MYZ}, \qquad C_{MYY} = \lambda_{XZ}C_{MXZ}, \qquad C_{MZZ} = \lambda_{XY}C_{MXY} \quad \textit{(Eq.27f)}$$

Figure 11 compares the square-plate normal-force coefficient $C_{FXX}$ with the numerical relation of Holmes et al. (2006) and the wind-tunnel data of Richards et al. (2008). The curves have a similar broad angular trend but do not agree pointwise. At the smallest nonzero plotted angle, the CFD coefficient is approximately four times the wind-tunnel value. Agreement improves over part of the intermediate range and worsens again at some orientations. The comparison therefore checks symmetry and order of magnitude; it is not a full validation. Plausible contributors include inflow turbulence, transition and edge separation, finite plate thickness, domain and boundary choices, and interpolation between simulated angles. The discrepancy propagates directly to trajectory and endpoint uncertainty.

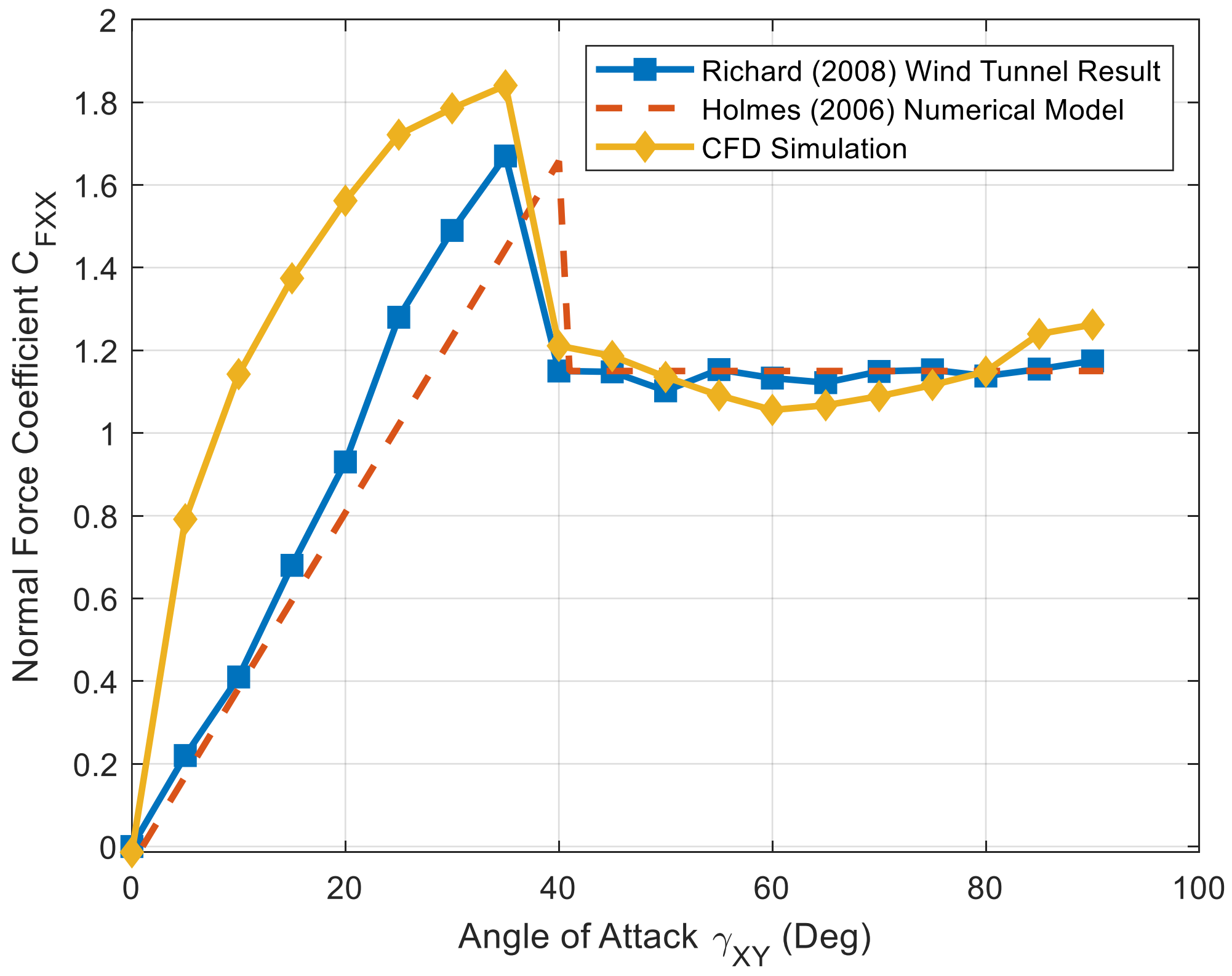

**Figure 11.** Normal-force coefficient from the planar CFD approximation compared with published numerical and wind-tunnel results. The low-angle discrepancy is material.

## 4 THREE-DIMENSIONAL NEAR-SURFACE WIND FIELD

A 3D numerical wind field is required to provide the wind speed information as the external force for the debris. Baker and Sterling Model is a notable three-dimensional vortex model, which has advantages in calculating the debris trajectories (Baker and Sterling, 2017). Baker and Sterling Model is designed under the simplified Euler Equations, which are derived from Navier-Stokes Equations (NSE) by neglecting the viscosity effects. The following simplification conditions are applied in the Baker and Sterling Model for computational convenience in the solution of Euler Equations listed as Eq. (28-31):

- **Steady mean field:** time derivatives are set to zero, $\partial / \partial t = 0$. Fluctuations and vortex wandering are not resolved.
- **Axisymmetry:** azimuthal derivatives are set to zero, $\partial / \partial \theta = 0$. Local asymmetry and subvortices are outside the model.
- **Negligible body force in the air-flow equations:** $F = 0$. Gravity remains in the debris equations.
- **Incompressibility:** air density is treated as constant over the modeled domain.

$$U\frac{\partial U}{\partial r} + W\frac{\partial U}{\partial z} - \frac{V^2}{r} = -\frac{1}{\rho}\frac{\partial P}{\partial r} \qquad \textit{(Eq.28)}$$

$$U\frac{\partial V}{\partial r} + W\frac{\partial V}{\partial z} + \frac{UV}{r} = 0 \qquad \textit{(Eq.29)}$$

$$U\frac{\partial W}{\partial r} + W\frac{\partial W}{\partial z} = -\frac{1}{\rho}\frac{\partial P}{\partial z} \qquad \textit{(Eq.30)}$$

$$\frac{\partial (Ur)}{r\partial r} + \frac{\partial W}{\partial z} = 0 \qquad \textit{(Eq.31)}$$

Baker and Sterling (2017) prescribed a bounded radial-velocity form that varies with radius and height. Eq. (32) expresses $U$ using the scales $U_{max}$, $R_{max}$, and $Z_{max}$. Substitution into continuity gives $W$ in Eq. (33). The tangential equation yields $V$ in Eq. (34).

$$\frac{U}{U_{max}} = -4\frac{\left(\frac{r}{R_{max}}\right)\left(\frac{z}{Z_{max}}\right)}{\left(1+\left(\frac{r}{R_{max}}\right)^2\right)\left(1+\left(\frac{z}{Z_{max}}\right)^2\right)} \qquad \textit{(Eq.32)}$$

$$\frac{W}{U_{max}} = 4\left(\frac{Z_{max}}{R_{max}}\right)\frac{\ln\left(1+\left(\frac{z}{Z_{max}}\right)^2\right)}{\left(1+\left(\frac{r}{R_{max}}\right)^2\right)^2} \qquad \textit{(Eq.33)}$$

$$\frac{V}{U_{max}} = 2.88\frac{V_{max}}{U_{max}}\left(\frac{r}{R_{max}}\right)\frac{ln\left(1+\left(\frac{z}{Z_{max}}\right)^2\right)}{1+\left(\frac{r}{R_{max}}\right)^2} \qquad \textit{(Eq.34)}$$

Two similarity parameters control the field shape: the scale ratio $\delta = Z_{max}/R_{max}$ and the swirl ratio $S = V_{max}/U_{max}$. A translation speed $V_T$ moves the vortex center. The candidate field is indexed by ($R_{max}$, $U_{max}$, $\delta$, $S$), together with translation. Eq. (35) is the vertical-to-tangential ratio at the reference point $r/R_{max} = z/Z_{max} = 1$; it should not be read as the absolute maximum vertical velocity over the entire field. Figures 12 and 13 show the dimensionless radial and vertical profiles for $S = 2$ and $\delta = 1$.

$$\frac{W(R_{max},Z_{max})}{V(R_{max},Z_{max})} = (\ln 2)\frac{\delta}{S} \qquad \textit{(Eq.35)}$$

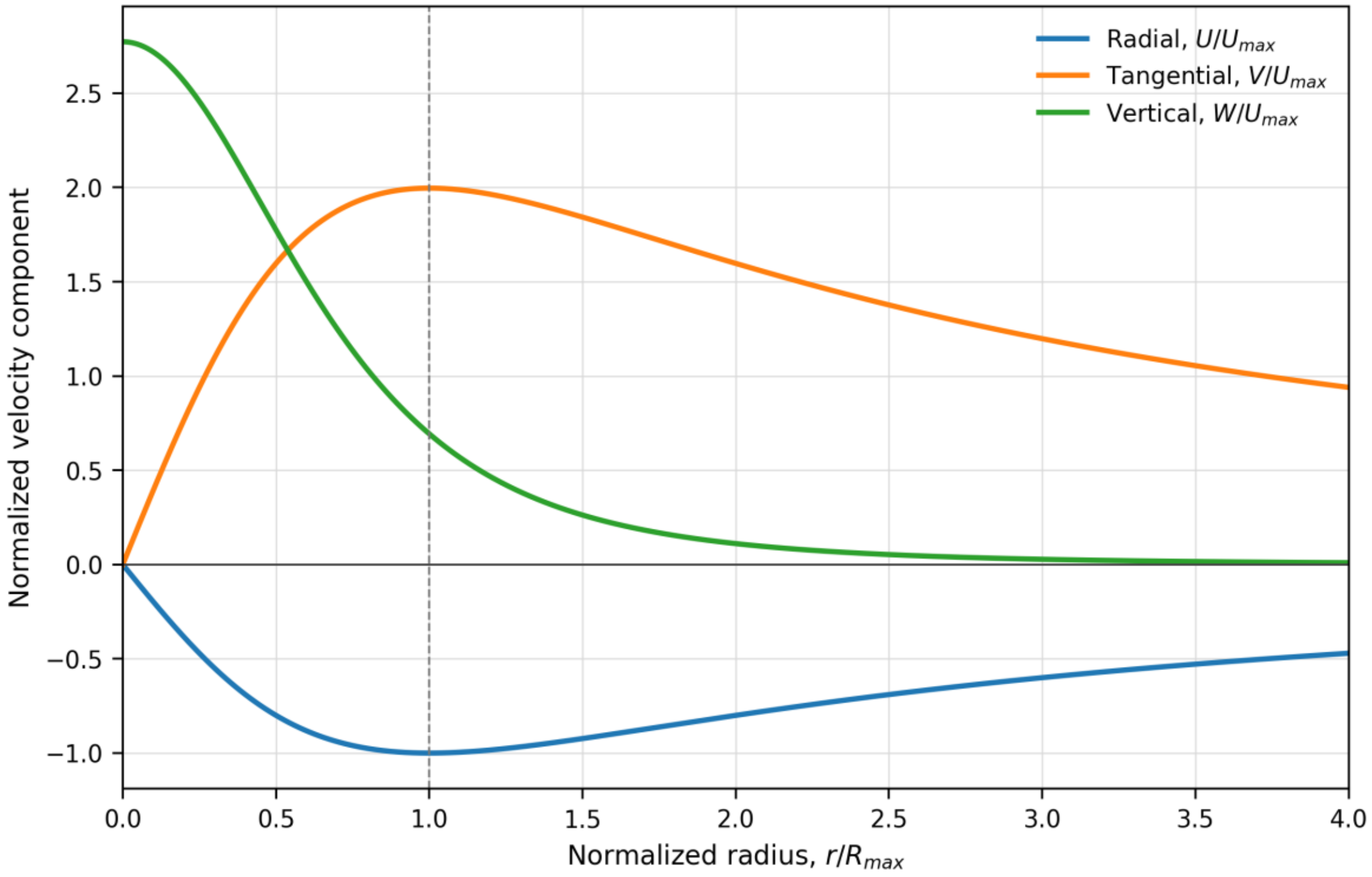

**Figure 12.** Dimensionless Baker-Sterling velocity components versus normalized radius at $z/Z_{max} = 1$, with $S = 2$ and $\delta = 1$. The dashed line marks $r/R_{max} = 1$.

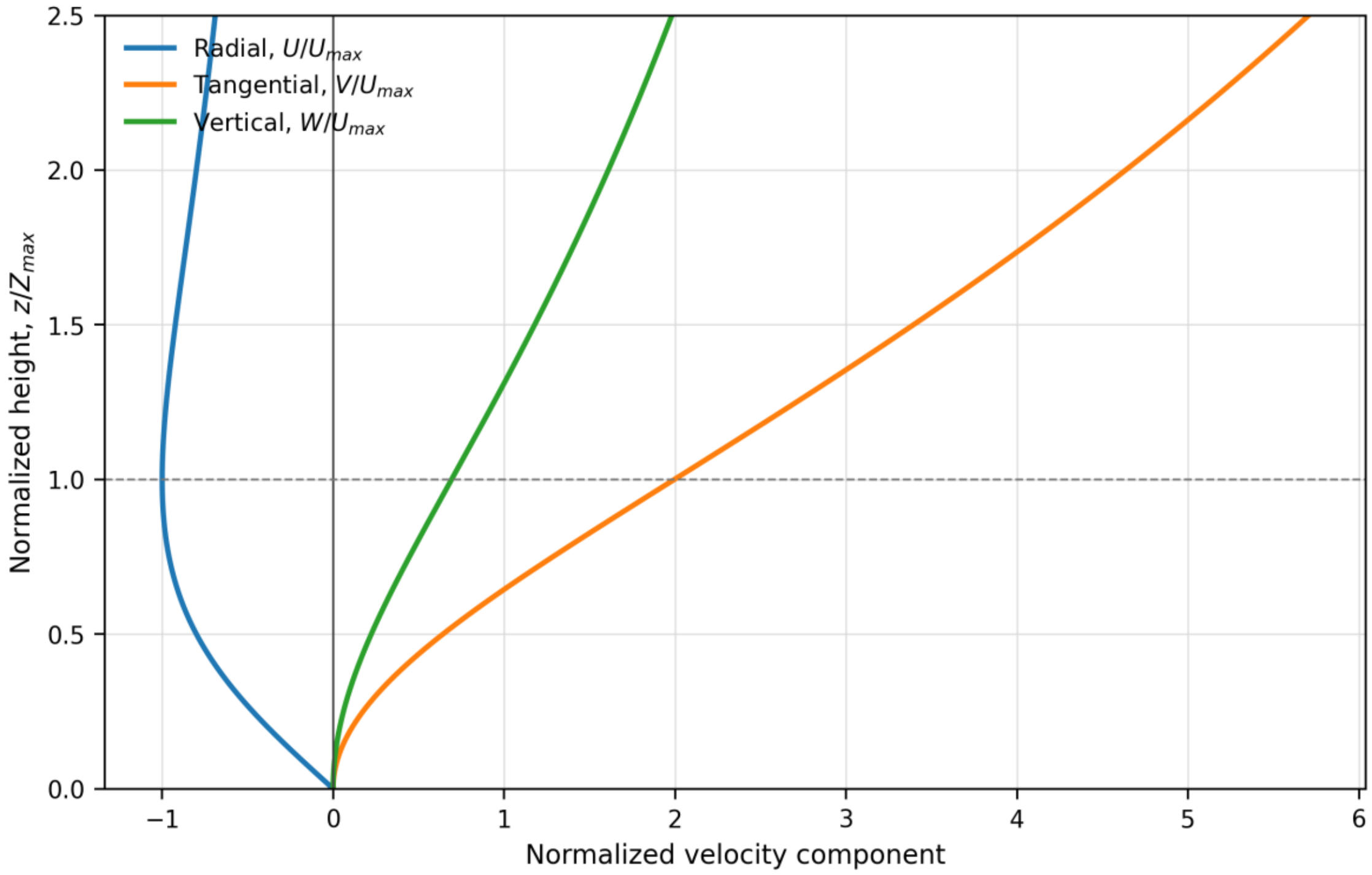

**Figure 13.** Dimensionless Baker-Sterling velocity components at $r/R_{max} = 1$, with velocity on the horizontal axis and normalized height on the vertical axis ($S = 2$, $\delta = 1$).

The profiles expose limitations relevant to inversion. The steady axisymmetric field does not reproduce tornado turbulence, wandering, local asymmetry, terrain effects, or building-vortex interaction. The vertical and tangential components also continue to increase over part of the plotted height range. The model is retained for computational efficiency, not as a complete representation of a turbulent tornado; fitted parameters may compensate for omitted physics.

## 5 TIME-MARCHING SOLUTION FOR DEBRIS TRAJECTORIES

After release, each roof panel is represented as a rigid rectangular body with three translational and three rotational degrees of freedom. The global debris centers ($X_{m,i}$, $Y_{m,i}$, $Z_{m,i}$) form a trajectory through the resolved wind field ($V_{X,i}$, $V_{Y,i}$, $V_{Z,i}$). Two right-handed Cartesian frames are used: a ground-fixed global frame $G = (X, Y, Z)$ and a body-fixed frame $B = (x_{\mathrm{b}}, y_{\mathrm{b}}, z_{\mathrm{b}})$ aligned with the panel's principal axes at its mass center.

The state at time $t_{\mathrm{i}}$ is $S_{\mathrm{i}} = \{\boldsymbol{r}_{\mathrm{m,i}}^{\mathrm{G}}, \boldsymbol{v}_{\mathrm{m,i}}^{\mathrm{G}}, \mathbf{R}_{\mathrm{BG,i}}, \boldsymbol{\omega}_{\mathrm{i}}^{\mathrm{B}}\}$, and it is advanced to $i + 1$ over $\Delta t$, Here, $\boldsymbol{r}_{\mathrm{m}}^{\mathrm{G}}$ and $\boldsymbol{v}_{\mathrm{m}}^{\mathrm{G}}$ are the mass-center position and velocity, $\mathbf{R}_{\mathrm{BG}}$ maps body-frame components to the global frame, and $\boldsymbol{\omega}^{\mathrm{B}}$ is the body angular-velocity vector. Consequently, $\boldsymbol{a}^{\mathrm{G}} = \mathbf{R}_{\mathrm{BG}}\boldsymbol{a}^{\mathrm{B}}$ and $\boldsymbol{a}^{\mathrm{B}} = {\mathbf{R}_{\mathrm{BG}}}^{\mathrm{T}}\boldsymbol{a}^{\mathrm{G}}$.

Then, the required initial conditions are:

- Vortex-center coordinates in the global frame: $(X_0, Y_0, Z_0)$.
- Debris mass-center coordinates in the global frame: $(X_{m,0}, Y_{m,0}, Z_{m,0})$.
- Debris translational velocity: $(\dot{X}_{m,0}, \dot{Y}_{m,0}, \dot{Z}_{m,0})$.
- Debris orientation expressed by Euler angles: $(\theta_{roll,0}, \theta_{pitch,0}, \theta_{yaw,0})$.
- Initial angular velocity or Euler-angle rates: $(\dot{\theta}_{roll,0}, \dot{\theta}_{pitch,0}, \dot{\theta}_{yaw,0})$.

At each integration substep, the debris position relative to the vortex center is expressed in cylindrical coordinates as Eq. (36). The $\tan^{-1}$ function preserves the azimuthal quadrant and remains defined when $X_{\mathrm{m,i}} = X_{\mathrm{c,i}}$. The Baker-Sterling vortex model supplies the radial, tangential, and vertical velocity components $U_{\mathrm{i}}$, $V_{\mathrm{i}}$, and $W_{\mathrm{i}}$.

$$\begin{bmatrix} R_{m,i} \\ \theta_{m,i} \\ Z_{m,i} \end{bmatrix} = \begin{bmatrix} \sqrt{(X_{m,i} - X_i)^2 + (Y_{m,i} - Y_i)^2} \\ \tan^{-1} \frac{Y_{m,i} - Y_i}{X_{m,i} - X_i} \\ Z_{m,i} - Z_i \end{bmatrix} \qquad (Eq.36)$$

With vortex translation velocity $\boldsymbol{V}_{\mathrm{T}}^{\mathrm{G}} = [V_{\mathrm{TX}} \quad V_{\mathrm{TY}} \quad 0]^{\mathrm{T}}$, the global wind velocity Eq. (37a), and debris-relative-air velocity are Eq. (37b).

$$\boldsymbol{u}_{\mathrm{w,i}}^{\mathrm{G}} = \begin{bmatrix} U_{\mathrm{i}}\cos(\alpha_{\mathrm{i}}) - V_{\mathrm{i}}\sin(\alpha_{\mathrm{i}}) \\ U_{\mathrm{i}}\sin(\alpha_{\mathrm{i}}) + V_{\mathrm{i}}\cos(\alpha_{\mathrm{i}}) \\ W_{\mathrm{i}} \end{bmatrix} + \boldsymbol{V}_{\mathrm{T}}^{\mathrm{G}} \qquad \textit{(Eq.37a)}$$

$$\boldsymbol{u}_{\mathrm{rel,i}}^{\mathrm{B}} = \mathbf{R}_{\mathrm{BG,i}}{}^{\mathrm{T}}\left(\boldsymbol{v}_{\mathrm{m,i}}^{\mathrm{G}} - \boldsymbol{u}_{\mathrm{w,i}}^{\mathrm{G}}\right) \qquad \textit{(Eq.37b)}$$

The sign in the second relation follows the adopted aerodynamic-coefficient convention: the coefficient functions return loads opposing the debris velocity relative to the air. Let $\boldsymbol{C}_{\mathrm{F}}$ and $\boldsymbol{C}_{\mathrm{M}}$ denote the plate force and moment coefficient vectors evaluated from $\boldsymbol{u}_{\mathrm{rel}}^{\mathrm{B}}$. The body-frame aerodynamic loads are Eq. (38).

$$\boldsymbol{F}_{\mathrm{a,i}}^{\mathrm{B}} = \frac{1}{2}\rho_{\mathrm{a}}\left\|\boldsymbol{u}_{\mathrm{rel,i}}^{\mathrm{B}}\right\|^{2} \mathrm{diag}(bc, ac, ab)\, \boldsymbol{C}_{\mathrm{F,i}}$$

*(Eq.38a)*

$$\boldsymbol{M}_{\mathrm{a,i}}^{\mathrm{B}} = \frac{1}{2}\rho_{\mathrm{a}}\left\|\boldsymbol{u}_{\mathrm{rel,i}}^{\mathrm{B}}\right\|^{2} \mathrm{diag}(ac^{2}, cb^{2}, bc^{2})\, \boldsymbol{C}_{\mathrm{M,i}} \qquad \textit{(Eq.38b)}$$

The diagonal scale matrices retain the coefficient normalization used by the aerodynamic model; $a$, $b$, and $c$ are the panel dimensions along its body axes. For material density $\rho_{\mathrm{m}}$, the mass and principal inertia tensor are Eq. (39):

$$m = \rho_{\mathrm{m}} abc, \qquad \mathbf{I}_{\mathrm{B}} = \frac{m}{12}\mathrm{diag}(b^{2} + c^{2}, a^{2} + c^{2}, a^{2} + b^{2})$$

*(Eq.39)*

Translation is evaluated in the global frame, with gravity directed along the negative global $Z$ axis in Eq. (40):

$$\dot{\boldsymbol{r}}_{\mathrm{m}}^{\mathrm{G}} = \boldsymbol{v}_{\mathrm{m}}^{\mathrm{G}} \qquad \textit{(Eq.40a)}$$

$$\dot{\boldsymbol{v}}_{\mathrm{m}}^{\mathrm{G}} = \frac{1}{m}\mathbf{R}_{\mathrm{BG}}\boldsymbol{F}_{\mathrm{a}}^{\mathrm{B}} + [0 \quad 0 \quad -g]^{\mathrm{T}} \qquad \textit{(Eq.40b)}$$

Rotational motion is evaluated in the body frame. The initial attitude follows the right-handed 3-2-1 (yaw-pitch-roll) convention $\mathbf{R}_{\mathrm{BG},0} = \mathbf{R}_{\mathrm{z}}(\psi_0)\mathbf{R}_{\mathrm{y}}(\theta_0)\mathbf{R}_{\mathrm{x}}(\varphi_0)$, and the governing rotational equations are Eq. (41)

$$\dot{\boldsymbol{\omega}}^{\mathrm{B}} = \mathbf{I}_{\mathrm{B}}{}^{-1}[\boldsymbol{M}_{\mathrm{a}}^{\mathrm{B}} - \boldsymbol{\omega}^{\mathrm{B}} \times (\mathbf{I}_{\mathrm{B}}\boldsymbol{\omega}^{\mathrm{B}})] \quad \textit{(Eq.41a)}$$

$$\dot{\mathbf{R}}_{\mathrm{BG}} = \mathbf{R}_{\mathrm{BG}}[\boldsymbol{\omega}^{\mathrm{B}}]_{cross} \quad \textit{(Eq.41b)}$$

Note: $[\boldsymbol{\omega}]_{cross} = \begin{bmatrix} 0 & -\omega_{\mathrm{z}} & \omega_{\mathrm{y}} \\ \omega_{\mathrm{z}} & 0 & -\omega_{\mathrm{x}} \\ -\omega_{\mathrm{y}} & \omega_{\mathrm{x}} & 0 \end{bmatrix}$

This rotation-matrix kinematics replaces the inverse Euler-rate mapping in the earlier formulation. It therefore remains regular as the attitude passes through a pitch of $\pm 90°$. For a uniformly translating vortex, the center advances as

$$\boldsymbol{r}_{\mathrm{c,i+1}}^{\mathrm{G}} = \boldsymbol{r}_{\mathrm{c,i}}^{\mathrm{G}} + \Delta t \boldsymbol{V}_{\mathrm{T}}^{\mathrm{G}} \quad \textit{(Eq.42)}$$

When a discrete vortex-center path is prescribed, its coordinates are linearly interpolated to the midpoint of each integration substep rather than held fixed over the full field time interval. Let $h = \frac{\Delta t}{n_{\mathrm{s}}}$ denote a translational substep. Aerodynamic loads are reevaluated at its start; velocity and position then advance according to Eq. (43):

$$\boldsymbol{v}_{\mathrm{m,j+1}}^{\mathrm{G}} = \boldsymbol{v}_{\mathrm{m,j}}^{\mathrm{G}} + h\boldsymbol{a}_{\mathrm{j}}^{\mathrm{G}} \quad \textit{(Eq.43a)}$$

$$\boldsymbol{r}_{\mathrm{m,j+1}}^{\mathrm{G}} = \boldsymbol{r}_{\mathrm{m,j}}^{\mathrm{G}} + \frac{h}{2}\left(\boldsymbol{v}_{\mathrm{m,j}}^{\mathrm{G}} + \boldsymbol{v}_{\mathrm{m,j+1}}^{\mathrm{G}}\right) \quad \textit{(Eq.43b)}$$

Acceleration is held constant only within the current substep. The body angular velocity is integrated with a fourth-order Runge-Kutta scheme. Defining Eq. (45a), the update over a rotational substep $h_{\mathrm{r}}$ is shown in Eq.(45b)

$$\boldsymbol{f}(\boldsymbol{\omega}) = \mathbf{I}_{\mathrm{B}}{}^{-1}[\boldsymbol{M}_{\mathrm{a}}^{\mathrm{B}} - \boldsymbol{\omega} \times (\mathbf{I}_{\mathrm{B}}\boldsymbol{\omega})] \quad \textit{(Eq.45a)}$$

$$
\begin{aligned}
\boldsymbol{k}_1 &= \boldsymbol{f}(\boldsymbol{\omega}_\mathrm{j}) \\
\boldsymbol{k}_2 &= \boldsymbol{f}\left(\boldsymbol{\omega}_\mathrm{j} + \frac{h_\mathrm{r}}{2}\boldsymbol{k}_1\right) \\
\boldsymbol{k}_3 &= \boldsymbol{f}\left(\boldsymbol{\omega}_\mathrm{j} + \frac{h_\mathrm{r}}{2}\boldsymbol{k}_2\right) \\
\boldsymbol{k}_4 &= \boldsymbol{f}(\boldsymbol{\omega}_\mathrm{j} + h_\mathrm{r}\boldsymbol{k}_3) \\
\boldsymbol{\omega}_{\mathrm{j}+1} &= \boldsymbol{\omega}_\mathrm{j} + \frac{h_\mathrm{r}}{6}(\boldsymbol{k}_1 + 2\boldsymbol{k}_2 + 2\boldsymbol{k}_3 + \boldsymbol{k}_4)
\end{aligned}
\qquad \textit{(Eq. 45b)}
$$

Then, the attitude is advanced directly on $\mathrm{SO}(3)$ with the midpoint body angular velocity in Eq.(46a). Here, $N\{\cdot\}$ denotes reorthonormalization after each finite rotation. For $\vartheta = \|\Delta\boldsymbol{\vartheta}\|$, the Rodrigues increment is Eq. (46b).

$$
\begin{aligned}
\Delta\boldsymbol{\vartheta}_\mathrm{j} &= \frac{h_\mathrm{r}}{2}(\boldsymbol{\omega}_\mathrm{j} + \boldsymbol{\omega}_{\mathrm{j}+1}) \\
\mathbf{R}_{\mathrm{BG,j+1}} &= N\left\{\mathbf{R}_{\mathrm{BG,j}} \exp\left([\Delta\boldsymbol{\vartheta}_\mathrm{j}]_{cross}\right)\right\}
\end{aligned}
\qquad \textit{(Eq. 46a)}
$$

$$
\exp([\Delta\boldsymbol{\vartheta}]_{cross}) = \mathbf{I} + \frac{\sin(\vartheta)}{\vartheta}[\Delta\boldsymbol{\vartheta}]_\times + \frac{1-\cos(\vartheta)}{\vartheta^2}[\Delta\boldsymbol{\vartheta}]_{cross}{}^2 \qquad \textit{(Eq. 46b)}
$$

The small-angle limit $\mathbf{I} + [\Delta\boldsymbol{\vartheta}]_\times$ is used as $\vartheta$ approaches zero. Reorthonormalization maintains $\mathbf{R}_{\mathrm{BG}}{}^\mathrm{T}\mathbf{R}_{\mathrm{BG}} = \mathbf{I}$ and $\det(\mathbf{R}_{\mathrm{BG}}) = 1$ without evaluating a singular Euler-rate transformation. Time marching continues while the panel remains airborne. When the mass-center height first crosses $Z = 0$, the impact time and horizontal landing point are linearly interpolated between the final airborne state and the first below-ground trial state. The resulting endpoint is retained for comparison with the unpaired observed debris distribution.

# 6 APPLICATION FOR NEAR-SURFACE TORNADIC FIELD RECONSTRUCTION

## 6.1 *Case definition and mapped evidence*

This application examines whether debris mapped around a damaged building can constrain a near-surface tornadic wind field. Figure 14 documents a target building affected by the 2011 Joplin, Missouri, tornado and the surrounding debris identified manually from post-event aerial imagery.

The source is treated as a two-story, single-family residence. Most mapped plate-like debris has an approximate 1:2 aspect ratio, consistent with the nominal roof-panel dimensions $(4ft \times 8ft \times 0.5\,inch)$ . The panels are modeled as plywood with mass density $\rho_m = 680\ kg/m^3$ and thickness $12.7\ mm$ (Vladimirova and Gong, 2024).

Figure 15 defines a $1km^2$ analysis region around the target building. The prescribed vortex center begins at ($X_0 = 0, Y_0 = 0, Z_0 = 0$) and translates along the saved path at $V_T = 13.4\ m/s$, taken from the Joplin damage investigation (Kuligowski et al., 2014), to ($X_0 = 1000, Y_0 = -87.5$ , $Z_0 = 0$) over duration $T$. The source building is represented by a 4:12 gable roof with an eave height of 5.5 m (18 ft). The roof is discretized into $N = 82$ candidate panels whose centroids define the initial debris mass centers. Eq. (47) gives the initial coordinates of panel $n$, where $n \in [1, N]$; $Y_{m,ridge,0}$ denotes the ridge-line coordinate. Eq. (48) specifies the initial panel orientation using the Euler angles $\theta_{roll,0}$, $\theta_{pitch,0}$, and $\theta_{yaw,0}$. Each panel is assumed to remain fixed before release; consequently, its initial 6-DOF translational and angular velocities are set to 0.

$$\begin{bmatrix} X_{m,0} \\ Y_{m,0} \\ Z_{m,0} \end{bmatrix} = \begin{bmatrix} X_{m,n,0} \\ Y_{m,n,0} \\ 5.5 + \frac{1}{3}\left(Y_{m,ridge,0} - Y_{m,n,0}\right) \end{bmatrix} \tag{Eq.47}$$

$$\begin{bmatrix} \theta_{yaw,0} \\ \theta_{pitch,0} \\ \theta_{roll,0} \end{bmatrix} = \begin{bmatrix} 0^{\circ} \\ 71.6^{\circ} \\ 0^{\circ} \end{bmatrix} \tag{Eq.48}$$

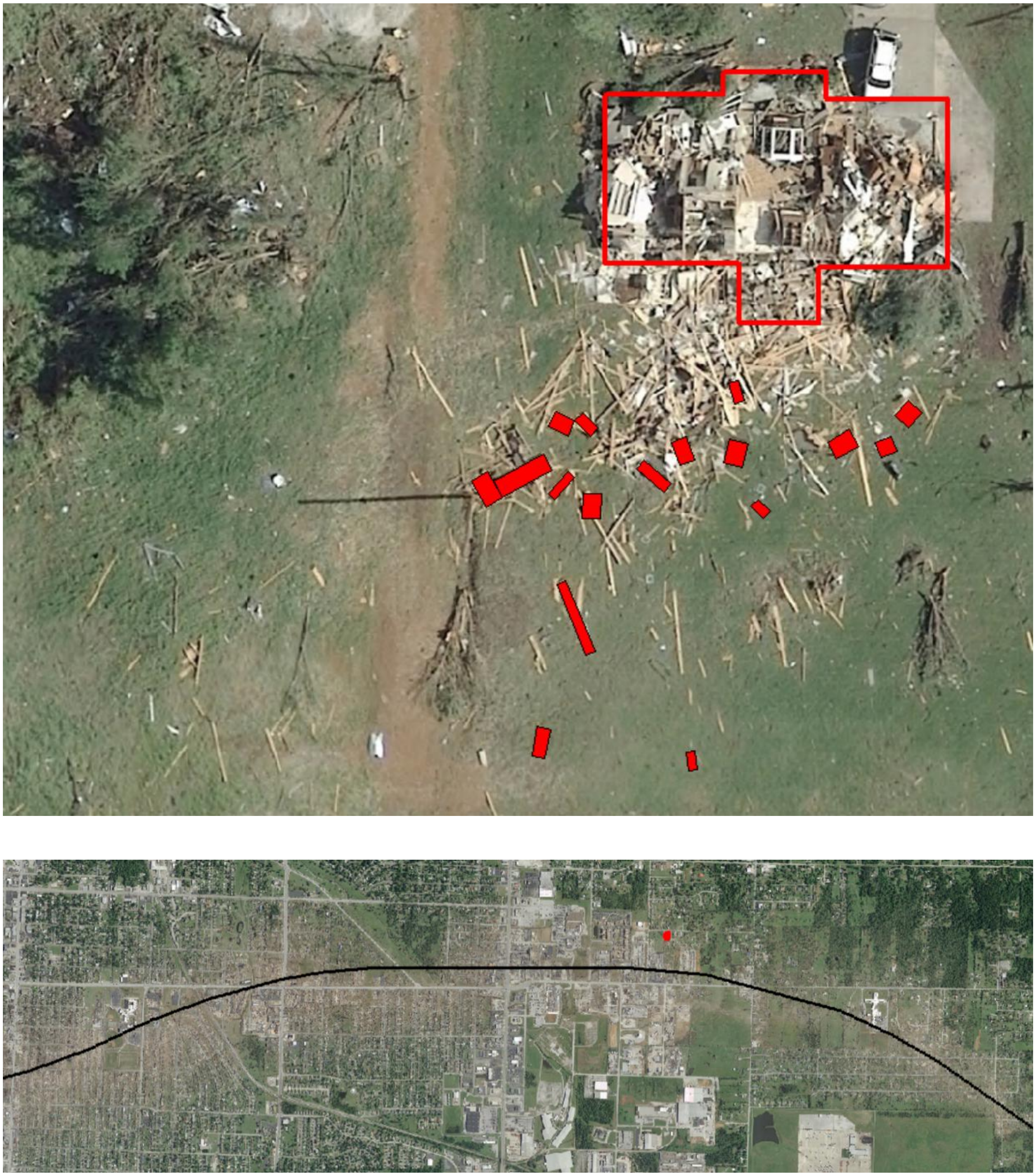

**Figure 14.** Post-event evidence used to define the Joplin application. (Top) Plate-like windborne debris (red markers) and the footprint of the damaged source building (red outline). (Bottom)

Relative positions of the selected debris region (red markers) and tornado path (black line) in Joplin, Missouri.

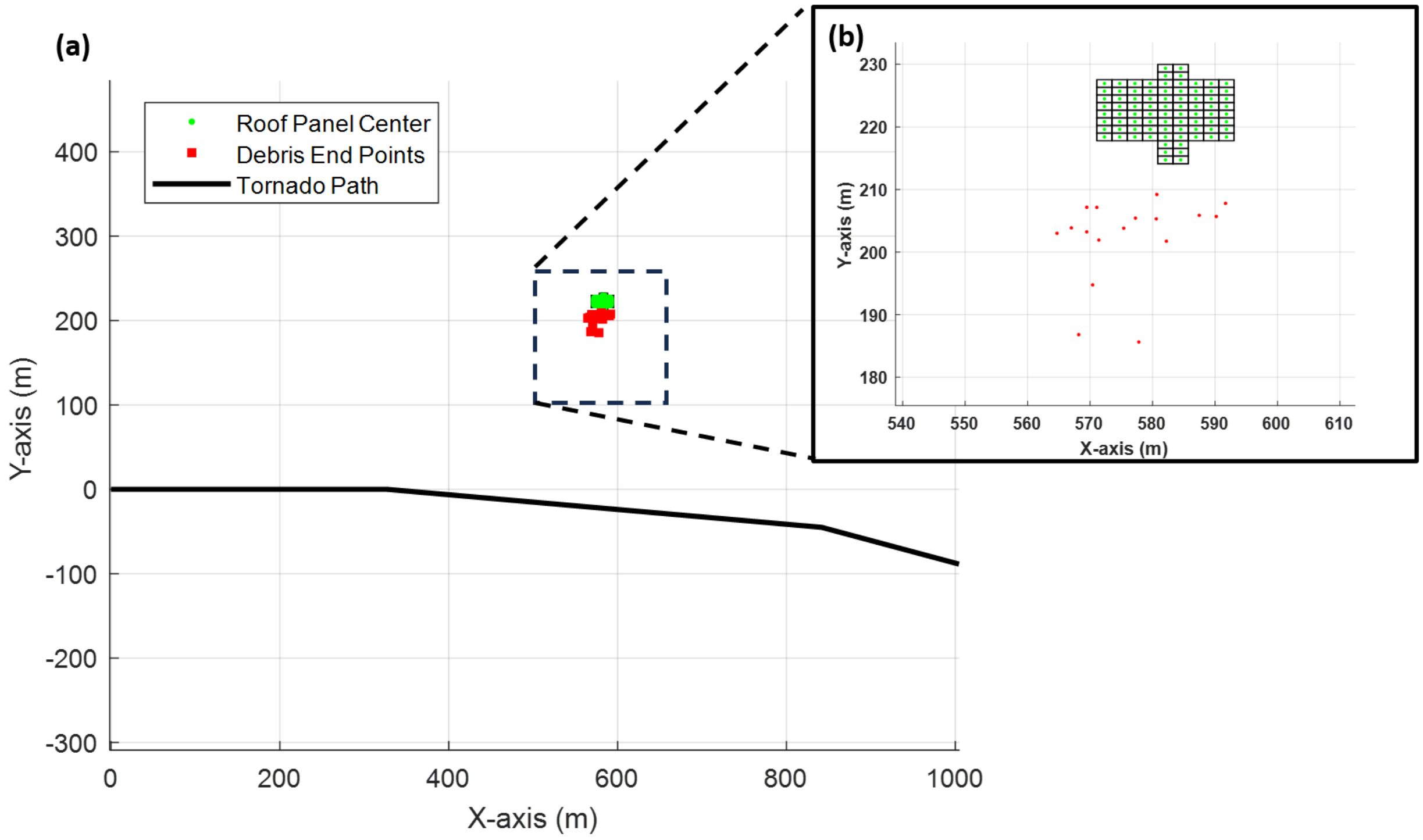

**Figure 15.** Prescribed Joplin reconstruction domain. Panel (a) shows the saved vortex-center path, candidate source roof, 17 mapped debris endpoints, and analysis boundary. Panel (b) classifies the 82 candidate source panels as interior, edge, or corner panels. No source-to-endpoint correspondence is assumed. Coordinates are expressed in the archived local planar system.

### 6.2 *Roof-panel release and fragility models*

Roof-panel release was modeled probabilistically rather than prescribed by a universal critical wind speed. The approximately 39 m/s threshold reported by Visscher and Kopp, (2007) is specific to their full-scale-equivalent roof-sheathing experiment and was not transferred directly to the present application. Instead, release is governed by the limit-state function in Eq. (49): a panel enters flight when $g(X) \leq 0$. Here, R denotes fastener resistance per unit area, W is the aerodynamic uplift-pressure magnitude, and D is the panel dead load.

The stored curves evaluated in this subsection comprise two distinct screening formulations. The first preserves the archived Lee-and-Rosowsky-based conventional-wind implementation and the project inputs in Table 1, including D = 84.7 N/m². It is not a numerical replication of Lee and Rosowsky (2005): the archived code uses Monte Carlo simulation rather than their first-order reliability method, retains the subtraction form (|GCp| - GCpi), and represents only the three full-size panel zones in the source grid; the half-size short-edge class considered by Lee and Rosowsky is omitted. The second formulation is an ASCE 7-22-informed tornado release screen. Its pressure coefficients and stochastic assumptions were specified before the debris endpoint analysis and were not calibrated to improve endpoint agreement.

$$g(X) = R - (W - D) \quad \text{(Eq.49)}$$

$$W = 0.613 K_z K_{zt} K_d V^2 I (GC_p - GC_{pi}) \quad \text{(Eq.50)}$$

**Table 1.** Legacy Lee-Rosowsky (2005) conventional-wind inputs used in the archived release screen

| **Panel location** | ***R*** **(N/m²)** | ***D*** **(N/m²)** | $K_z$ | $K_d$ | $GC_p$ | $GC_{pi}$ |
|---|---|---|---|---|---|---|
| **Edge** | 2760 (0.20) | 84.7 (0.10) | 0.71 (0.19) | 0.89 (0.16) | -1.455 (0.12) | 0.15 (0.33) |
| **Corner** | 2760 (0.20) | 84.7 (0.10) | 0.71 (0.19) | 0.89 (0.16) | -1.768 (0.12) | 0.15 (0.33) |
| **Interior** | 2760 (0.20) | 84.7 (0.10) | 0.71 (0.19) | 0.89 (0.16) | -0.855 (0.12) | 0.15 (0.33) |

*Note 1: Values are means; coefficients of variation (COVs) are given in parentheses.*

*Note 2: Fastener resistance corresponds to 8d common nails (3.33-mm diameter and 63.5-mm length), following the archived Lee-based implementation.*

*Note 3: The archived dead load, D = 84.7 N/m², represents 12.7-mm plywood with an assumed density of 680 kg/m³; it should not be interpreted as an exact reproduction of every Lee and Rosowsky (2005) benchmark case.*

*Note 4: The GCp entries are roof-suction magnitudes. Kz and Kd are conventional-wind factors retained solely to document the archived calculation.*

The tornado sensitivity follows the ASCE 7-22 pressure framework as a research screening implementation, with interpretation guided by the FEMA-NIST design guide (ASCE, 2022; FEMA and NIST, 2023). Coefficient attribution was checked against the ASCE 7-22 supplements and errata available from ASCE as of July 2026 (ASCE, 2026). Because external roof suction is negative by sign convention, Eq. (51) expresses the corresponding uplift demand as a positive magnitude.

$$W = 0.613\, K_{zTor}\, K_e V_T^2 \left[ K_{dT} K_{vT}\, |GC_p| \,+\, GC_{piT} \right] \qquad \text{(Eq.51)}$$

The inverse calculation adopts the ordinary-building component-and-cladding directionality factor $K_{dT} = 0.75$. For the 4:12 roof slope (18.4°), nominal $GC_p$ values are taken from ASCE 7-22 for a 7°-20° gable and the minimum effective-wind-area branch at or below 10 ft² (0.93 m²). Although each sheet has a gross area of 32.04 ft² (2.98 m²), the release screen treats the effective wind area as the connection or fastener tributary area, approximately 1-2 ft² in the Lee model, because R is a connection-level resistance. This interpretation is a screening assumption. The archived three-zone map does not resolve Zone 1′, and a panel-wide effective-area interpretation would require recalculated coefficients. The remaining nominal inputs are $K_{zTor} = 1.0$ at the low-rise roof height, the governing positive $GC_{piT} = +0.55$, and the $K_{vT}$ values from ASCE 7-22. Ground elevation is unavailable, so $K_e = 1.0$ is also an explicit screening assumption.

Table 2 lists the fixed coefficients. ASCE 7-22 supplies nominal design coefficients, not stochastic distributions. The calculation treats nominal $GC_p$ and $GC_{piT}$ values as distribution centers and assigns Lee and Rosowsky (2005) coefficients of variation; no nominal-to-mean bias conversion is applied. The resistance, dead-load, and coefficient distributions are research assumptions rather than uncertainty models specified by ASCE.

**Table 2.** Fixed ASCE 7-22-informed ordinary-building tornado release-screen parameters

| **Panel location** | $R$ **(N/m²)** | $D$ **(N/m²)** | $K_e$ | $K_{zTor}$ | $K_{dT}$ | $K_{vT}$ | $GC_p$ | $GC_{piT}$ |
|---|---|---|---|---|---|---|---|---|
| **Edge** | 2760 (0.20) | 84.7 (0.10) | 1.00 | 1.00 | 0.75 | 1.20 | −2.70 (0.12) | +0.55 (0.33) |
| **Corner** | 2760 (0.20) | 84.7 (0.10) | 1.00 | 1.00 | 0.75 | 1.20 | −3.60 (0.12) | +0.55 (0.33) |
| **Interior** | 2760 (0.20) | 84.7 (0.10) | 1.00 | 1.00 | 0.75 | 1.20 | −2.00 (0.12) | +0.55 (0.33) |

*Note: Parenthetical values on $GC_p$ and $GC_{piT}$ are research COV assumptions; nominal coefficients are used as distribution centers, and ASCE/SEI 7-22 does not specify these probability distributions.*

Each empirical fragility curve was represented by the fitted lognormal model in Eqs. (52) and (53), where $V_{50}$ is the median release speed, $\mu_{\ln V}$ is the mean of $\ln(V)$, $\beta$ is the fitted logarithmic dispersion, and $E[V]$ is the arithmetic mean release speed.

$$P(V) = \Phi\left[\frac{\ln(V) - \ln(V_{50})}{\beta}\right] \quad \text{Eq. (52)}$$

$$E[V] = \exp\left(\mu_{\ln V} + \frac{\beta^2}{2}\right), \qquad V_{50} = \exp(\mu_{\ln V}). \quad \text{Eq. (53)}$$

In Table 3, the legacy Lee (2005) based dataset uses 10,000 Monte Carlo samples and gives median release speeds of 76.39, 68.06, and 99.66 m/s for edge, corner, and interior panels, with logarithmic dispersions of 0.19022, 0.19343, and 0.18521, respectively. The ASCE 7-22-informed screen uses 500,000 samples and gives corresponding medians of 35.98, 30.55, and 40.89 m/s, with dispersions of 0.11733, 0.11735, and 0.11761. In both formulations, the corner panel is the most vulnerable, followed by the edge and interior panels. The lower tornado-screen medians follow from the adopted pressure, vertical-velocity, and internal-pressure terms; they do not constitute empirical validation of tornado-induced roof-panel release. For the tornado screen, all three conditional release probabilities are effectively unity at 78 m/s.

**Table 3.** Fitted lognormal parameters for the legacy Lee-based and ASCE 7-22-informed tornado release screens

| Model | Panel | $V_{50}$ median (m/s) | Arithmetic mean $E[V]$ (m/s) | $\mu_{\ln V}$ | Logarithmic dispersion $\beta$ |
|---|---|---|---|---|---|
| **Legacy Lee (2005) based** | Edge | 76.3877 | 77.7822 | 4.33582 | 0.19022 |
| **Legacy Lee (2005) based** | Corner | 68.0610 | 69.3462 | 4.22040 | 0.19343 |

| Model | Panel | $V_{50}$ median (m/s) | Arithmetic mean $E[V]$ (m/s) | $\mu_{\ln V}$ | Logarithmic dispersion $\beta$ |
|---|---|---|---|---|---|
| **Legacy Lee (2005) based** | Interior | 99.6603 | 101.3843 | 4.60177 | 0.18521 |
| **ASCE 7-22-informed** | Edge | 35.9815 | 36.2300 | 3.58301 | 0.11733 |
| **ASCE 7-22-informed** | Corner | 30.5495 | 30.7606 | 3.41935 | 0.11735 |
| **ASCE 7-22-informed** | Interior | 40.8856 | 41.1693 | 3.71078 | 0.11761 |

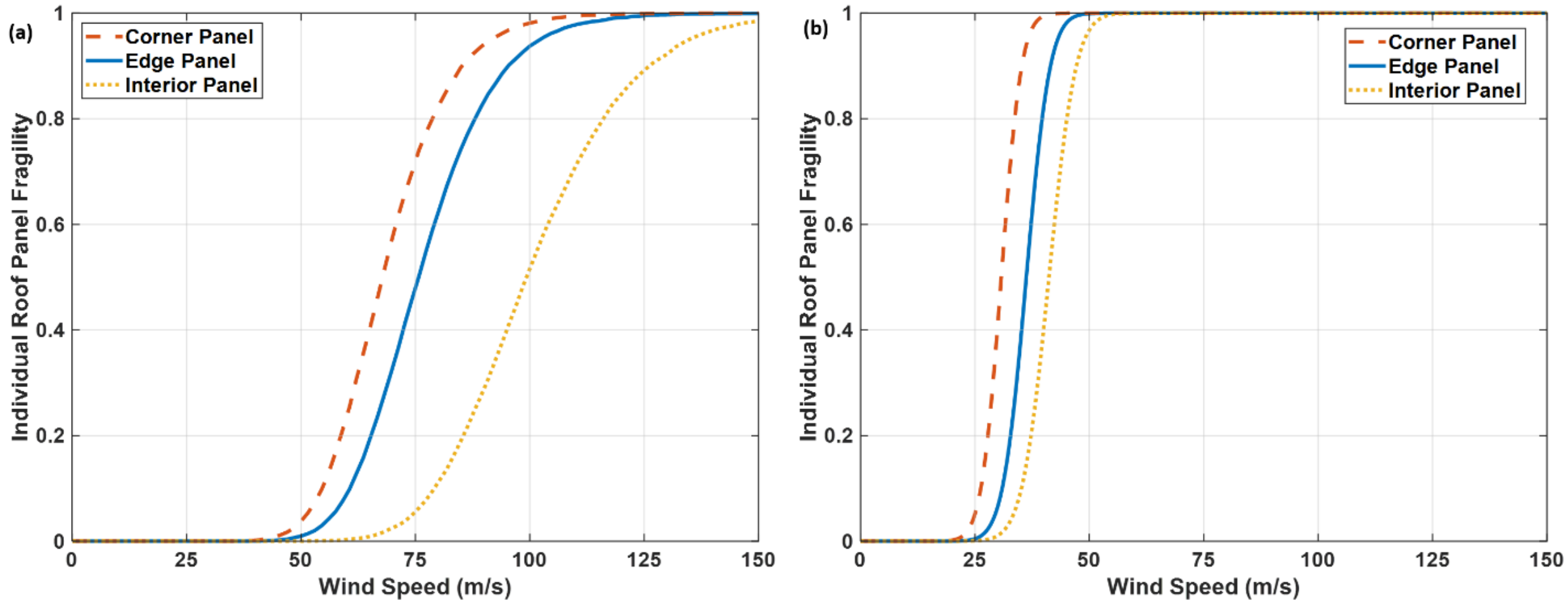

**Figure 16.** Individual roof-panel fragility curves for (a) the legacy Lee-based conventional-wind screen generated with 10,000 Monte Carlo samples and D = 84.7 N/m² and (b) the ASCE/SEI 7-

22-informed tornado screen generated with 500,000 samples. Solid, dashed, and dotted curves denote edge, corner, and interior panels, respectively. The curves are model-based screening results rather than empirically validated tornado release fragilities.

### 6.3 *Forward model and unequal-count point-set objective*

For a candidate Baker-Sterling parameter vector $\boldsymbol{\eta} = (R_{max}, U_{max}, \delta, S)$, the wind history at every roof panel is evaluated along the complete saved path. Here $U_{max}$ is the maximum radial-inflow scale, $S = V_{max}/U_{max}$ is the swirl ratio, $V_{max}$ is the tangential-velocity scale at the model reference location, and $\delta = z_{max}/R_{max}$. When the local wind exceeds a panel's fixed fragility threshold for the earliest time, the panel enters the six-degree-of-freedom flight calculation. Relative velocity is evaluated by vector subtraction in the global frame and then transformed to the body frame. Each released panel is integrated until its downward crossing of z=0; linear interpolation within the crossing step supplies the landing coordinates. Panels that do not release, remain airborne when the time period T ends, or encounter a numerical failure are not labeled or scored as landings.

Let $\mathcal{S}_{\boldsymbol{\eta}} = \{\mathbf{s}_i\}_{i=1}^{N_{\boldsymbol{\eta}}}$ be the simulated set and $\mathcal{O} = \{\mathbf{o}_j\}_{j=1}^{M}$ the observed set, with $M = 17$. The sets are unpaired; 82 candidate source panels cannot be equated one-to-one with 17 observations. The saved calculation used the mean distance from every simulated point to every observed point, Eq. (54). This legacy objective is not a one-to-one assignment and tends to favor the center of the observed cloud. Because object identities are unavailable, the reconstruction uses the bidirectional nearest-neighbor discrepancy in Eq. (55), together with sensitivity to missing and spurious observations.

$$E_{legacy}(\boldsymbol{\eta}) = \frac{1}{N_{\boldsymbol{\eta}} M} \sum_{i=1}^{N_{\boldsymbol{\eta}}} \sum_{j=1}^{M} \| \mathbf{s}_i - \mathbf{o}_j \|_2 \qquad \textit{(Eq.54)}$$

$$E_{sym}(\boldsymbol{\eta}) = \frac{1}{2N_{\boldsymbol{\eta}}} \sum_{i=1}^{N_{\boldsymbol{\eta}}} \min_j \| \mathbf{s}_i - \mathbf{o}_j \|_2 + \frac{1}{2M} \sum_{j=1}^{M} \min_i \| \mathbf{o}_j - \mathbf{s}_i \|_2 \qquad \textit{(Eq.55)}$$

The score has units of metres. The simulated-to-observed term penalizes simulated landings unsupported by the mapped cloud; the observed-to-simulated term penalizes observed regions not covered by a simulation. An empty simulated set receives an infinite score. The metric does not impose one-to-one correspondence or equal cardinality, and no count-matching penalty is used. Under the source-attribution assumption used to define this application, however, the 17 mapped items are a detected subset of actual source landings. The primary screen therefore requires $N_{\boldsymbol{\eta}} \geq 0.85\, N = 70$ while allowing any larger simulated count. A localization-only sensitivity relaxes this lower bound to show how sparse simulations can reduce the geometric score. The archived diagnostics include both directed terms, the simulated-set size $N_{\boldsymbol{\eta}}$ , release and airborne counts, and numerical status for every field. This formulation replaces the archived all-pairs sum, which concentrated simulated points near the observed-cloud center and treated stationary and airborne panels as endpoints.

## 6.4 *Reproducible parameter search and revised endpoint result*

The evaluated grid is listed in Table 4 for 20736 parameter combinations. Every candidate is evaluated at the saved $\Delta T = 0.012s$ path increment for tornado-like vortex translation; aerodynamic forces and translation are advanced with four 0.003 s substeps, and rotation is advanced with a fourth-order rigid-body-rate update. All candidates use the same panel-capacity quantiles and numerical settings. A candidate is ineligible when it produces fewer than 70 grounded panels or more than 5% numerically invalid released-panel trajectories. The complete score table retains eligible and rejected rows. The reported field is the minimum among count-compatible, numerically eligible candidates, not a claim of a unique physical optimum.

**Table 4. Executed reconstruction search grid**

| Parameter | Minimum | Maximum | Increment | Count |
|---|---|---|---|---|
| $\boldsymbol{R_{max}(m)}$ | 100 | 320 | 20 | 12 |
| $\boldsymbol{U_{max}(m/s)}$ | 35 | 90 | 5 | 12 |
| **S** | 0.4 | 1.5 | 0.1 | 12 |
| **δ** | 0.01 | 0.12 | 0.01 | 12 |

Figure 17 compares endpoint simulations generated with the two fragility formulations shown in Figure 16; within each case, every panel uses the fragility curve associated with its roof location. The unconstrained localization-only minimum was not retained because it produced too few grounded panels, even though its geometric score was lower. The application therefore retains the stated count screen: cases with $N_{\eta} < 0.85\, N$ landed panels are excluded. A small score obtained from only a few simulated contacts represents a sparse-set degeneracy rather than an acceptable primary reconstruction.

For each count-compatible candidate, $E_{sym}$ quantifies endpoint discrepancy. Eqs. (56)-(59), evaluated from Eqs. (32)-(35), give the near-surface resultant wind speed at the 5.5 m (18 ft) eave height and include vortex translation. Reporting the evaluation height, radial location, and reference frame resolves the earlier ambiguity in the term near-surface wind speed $V_{surface}$ between a 91 m/s component scale and a 76.05 m/s local resultant evaluated at $\mathrm{r} = R_{max}$ and $\mathrm{z} = 5.5\, m$.

$$U(\mathbf{z}, R_{max}) = -4U_{max}\left(\frac{z}{2Zmax+\frac{2z^2}{Zmax}}\right) \quad (Eq.56)$$

$$W(\text{z}, R_{max}) = U_{max}\ \delta\ln\left(1+\left(\frac{z}{Zmax}\right)^2\right) \quad (Eq.57)$$

$$V(\text{z}, R_{max}) = 1.44SU_{max}\ln\left(1+\left(\frac{z}{Zmax}\right)^2\right) \quad (Eq.58)$$

$$V_{surface}(\text{z}, R_{max}) = \left[U(\text{z}, R_{max})^2 + W(\text{z}, R_{max})^2 + V(\text{z}, R_{max})^2 + {V_T}^2\right]^{0.5} \quad (Eq.59)$$

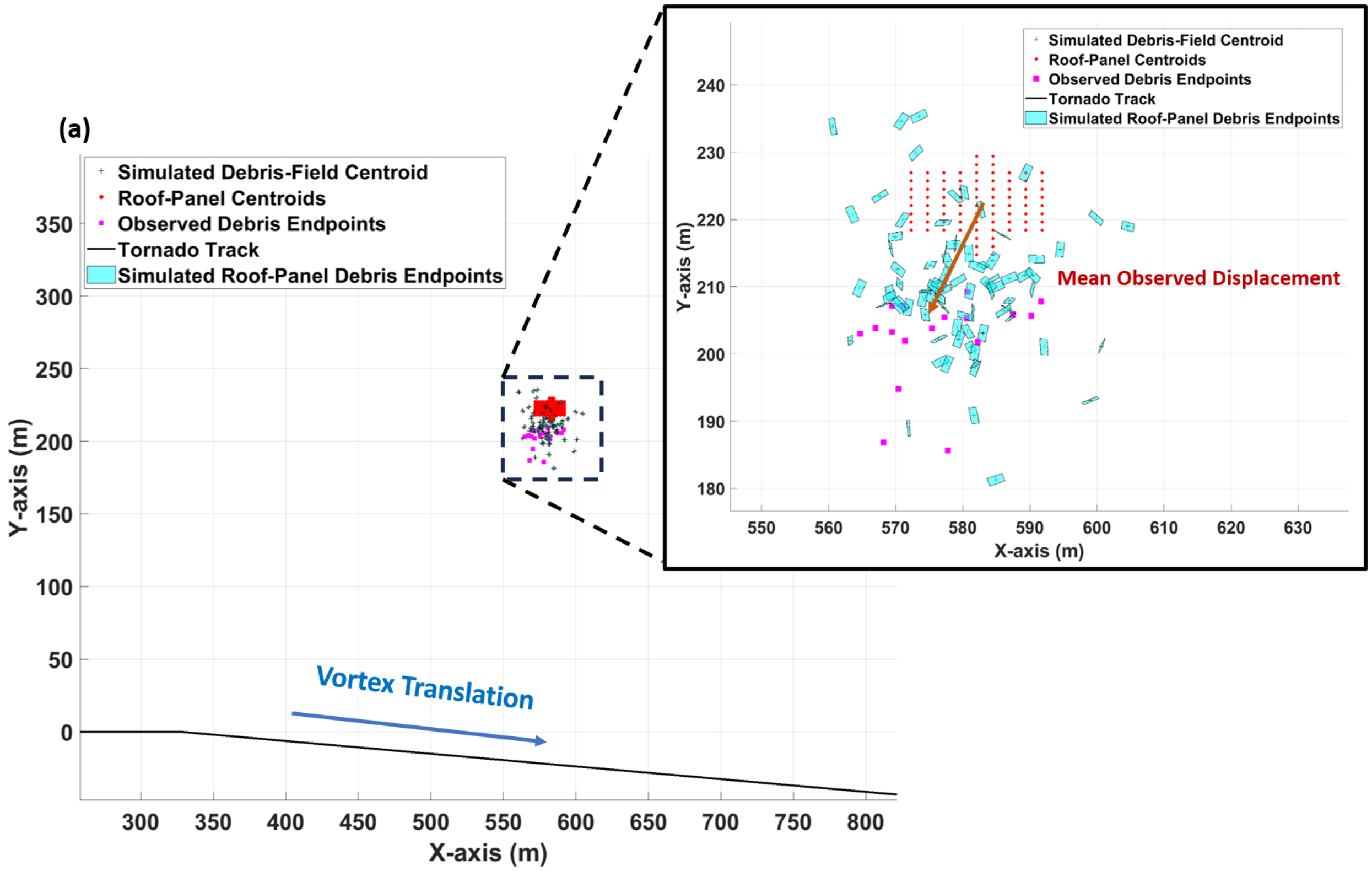

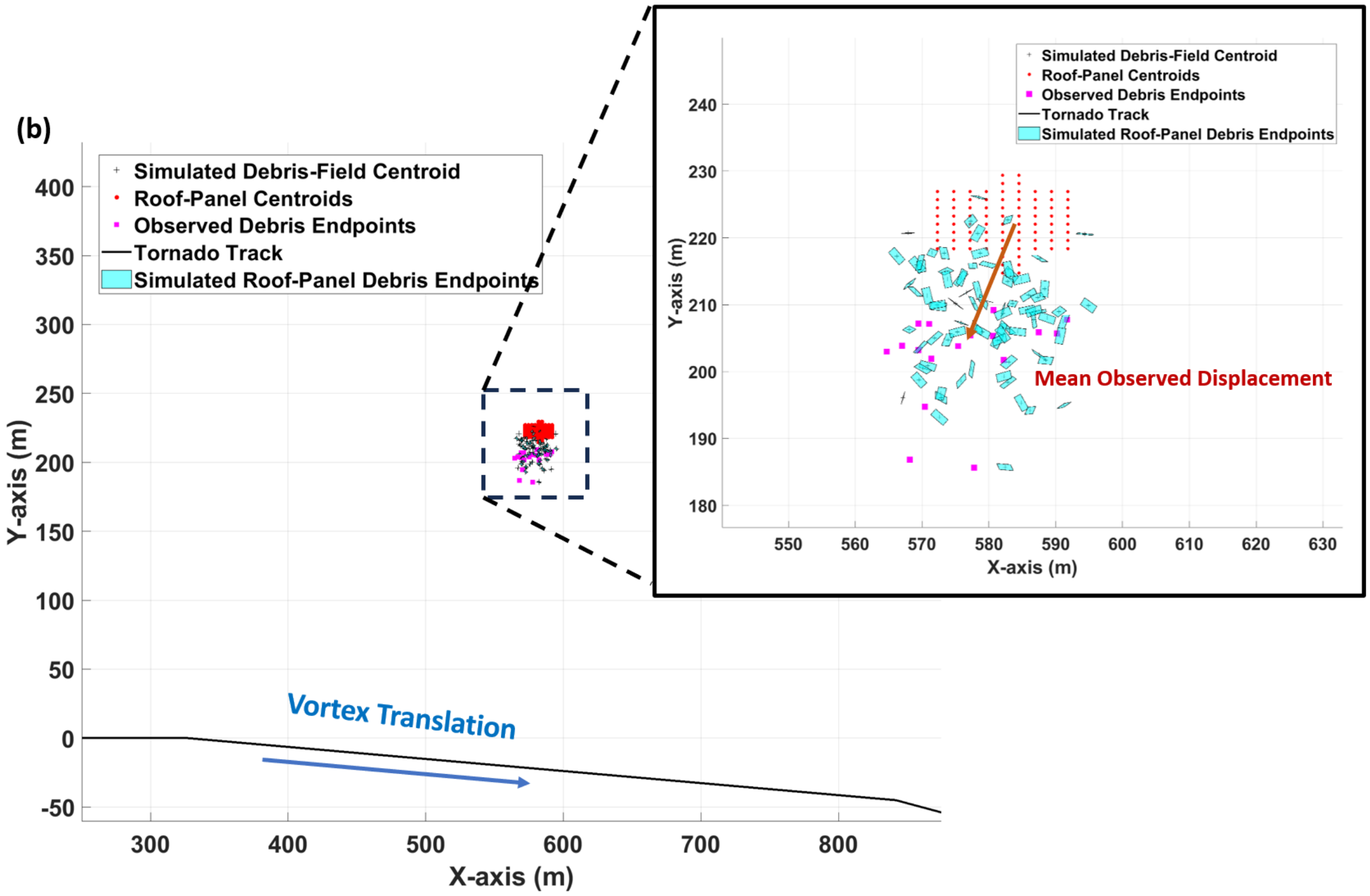

**Figure 17.** Simulated and observed endpoint distributions based on (a) the Lee and Rosowsky (2005) roof-panel fragilities and parameter set ($R_{max} = 200$ m, $U_{max} = 70$ m/s, $S = 0.9$, $\delta = 0.02$), for which $V_{surface}(5.5, 200) = 117.85\ m/s$.

(b) The ASCE 7-22-informed tornado fragilities and parameter set ($R_{max} = 280$ m, $U_{max} = 65$ m/s, $S = 1.2$, $\delta = 0.03$), for which $V_{surface}(5.5, 260) = 73.04\ m/s$. In both panels, magenta squares denote the unpaired observed endpoints, and teal outlines denote simulated roof-panel debris that reached the ground.

# 7 DISCUSSION AND LIMITATIONS

The fragility comparison explains why the revised release history differs materially from the legacy run. Under the original conventional-wind assumptions, the reproduced Lee and Rosowsky curves yield median release speeds of 68-99 m/s across the three full-size panel locations. The ASCE7-22 equivalent screen shifts those medians to 31-41 m/s because it uses tornado-specific pressure and vertical-velocity factors and a larger governing internal-pressure coefficient. These differences follow from the selected coefficients and uncertainty assumptions; they do not validate the tornado curves. Building-vortex interaction, progressive opening, shared framing, spatially correlated capacity, and pressure redistribution after initial component failure remain outside the release model.

The screened radius should not be characterized as close to the Rankine-vortex Rmax reported from Joplin tree-fall patterns by Chen and Lombardo (2019). The two estimates depend on different physical indicators, vortex families, objective functions, and spatial samples, and the debris fit remains poor. Their difference is therefore a cross-indicator comparison rather than an independent validation test.

**Table 5. Joplin wind-field comparison used to define the admissible radius and horizontal-speed ranges**

| Source | Indicator | $R_{max}$ (m) | Maximum Wind Speed (m/s) |
|---|---|---|---|
| **Present Study: Lee (2005)-Based Debris-Release Fragility Model** | Windborne Debris | 200 | 117.85 |
| **Present Study: ASCE 7-22-Informed Tornado Debris-Release Fragility Model** | Windborne Debris | 280 | 73.04 |
| **Karstens et al. (2013)** | Trees | 300 | 104 |
| **Kuligowski et al. (2014)** | Residential buildings and trees | 258 | 78.2 ± 15.7 |
| **Roueche et al. (2015)** | Residential buildings | 209–257 | 77.8 |
| **Lombardo et al. (2015)** | Trees | 257 | 78 |
| **Luo et al. (2015)** | Residential buildings | 220 | 71.2 ± 22.8 |
| **Chen and Lombardo (2019)** | Trees | 380 | 82.1 |

*Note: Values reproduce the supplied comparison table. Independent studies span Rmax = 209-300 m and maximum wind speed = 71.2-104 m/s. The Chen and Lombardo (2019) 380-m row is retained for comparison; Rankine-specific shape parameters are not transferred to the Baker-Sterling model.*

Four groups of uncertainty bound interpretation of the result. Observation uncertainty arises from manual debris classification, geolocation, occlusion, survey timing, and an unknown detection fraction. Release uncertainty includes fastener resistance, tributary area, roof-zone mapping, internal pressure, progressive failure, and the use of design coefficients as stochastic inputs. Trajectory uncertainty includes initial posture, quasi-steady aerodynamic coefficients, rotation, neglected turbulence and vortex wandering, time-step sensitivity, and the treatment of panels that leave the modeled record. Inverse and model-form uncertainty follows from the steady, axisymmetric Baker-Sterling family, prescribed path, sparse unpaired endpoints, and a Chamfer-type metric that can be insensitive to duplicated points. Consequently, different candidate fields can produce similar landing clouds.

Only one case is considered here. Future work should extend the analysis to multiple events and develop image-processing and machine-learning methods for recognizing debris in orthophotos. A larger case set would permit out-of-sample evaluation and more defensible characterization of uncertainty; it should not be assumed, by itself, to guarantee more accurate wind-field estimates.

The ASCE 7-22 and FEMA-NIST provisions used here are design guidance, not an inverse-model validation data set. Project design must follow the applicable ASCE 7-22 requirements for risk category, tornado speed, enclosure, effective area, directionality, and load combinations. Tornado shelters and safe rooms remain governed by guidance such as ICC 500 and FEMA P-361, not by this forensic calculation. A practical research program should (i) quantify endpoint geolocation and detection error; (ii) propagate building-level and panel-level release correlation; (iii) repeat the search with independent fragility seeds, alternative robust point-set metrics, and time steps; (iv) map a near-optimal parameter region rather than only a minimum; (v) test synthetic cases with known parameters; and (vi) compare independent events with radar, tree-fall, pressure, or trajectory observations.

## 8 CONCLUSIONS

This study formulates mapped windborne debris as an inverse point-set problem. A known source building, probabilistic release model, 6-DOF plate dynamics, and translating parametric vortex provide a forward map from candidate wind fields to static landing clouds. In the Joplin application, the reported parameter sets place simulated panels within the region occupied by the 17 mapped endpoints. The result supports the feasibility of using endpoint evidence as a forensic constraint, but it neither identifies a unique wind field nor independently validates a wind speed.

The next stage should emphasize independent reproducibility and validation: verify roof geometry and release orientation; verify and document the ASCE 7-22 tornado load terms; audit the point-set objective; independently repeat the full grid or a justified adaptive search; and report near-optimal parameter families under observation and model uncertainty. Wind-tunnel or higher-fidelity CFD data are also needed at the low angles where the present coefficients diverge most strongly. Until these checks are completed, debris-endpoint evidence should be treated as complementary to radar, tree-fall, and conventional damage indicators rather than as a stand-alone estimator.

## 9 AUTHOR CONTRIBUTIONS (CREDIT)

Guangzhao Chen: Conceptualization, methodology, software, formal analysis, visualization, and writing - original draft.

Franklin T. Lombardo: Conceptualization, resources, funding acquisition, supervision, and writing - review and editing.

## 10 ACKNOWLEDGMENTS

This work was supported by the U.S. National Science Foundation under Award No. 2053935, *Reconstruction of Four-Dimensional Near-Surface Wind Characteristics from Debris and Damage Attributes Using Computer Vision*.

## REFERENCES

Abdelhady, A.U., Spence, S.M.J., McCormick, J., 2021. A three-dimensional six-degree-of-freedom windborne debris trajectory model for tornadoes. J. Wind Eng. Ind. Aerodyn. 214, 104622. https://doi.org/10.1016/j.jweia.2021.104622

American Society of Civil Engineers (ASCE), 2022. Minimum Design Loads and Associated Criteria for Buildings and Other Structures, ASCE/SEI 7-22. Reston, VA. https://doi.org/10.1061/9780784415788

Baker, C.J., 2007. Wind engineering—Past, present and future. J. Wind Eng. Ind. Aerodyn. 95, 843–870. https://doi.org/10.1016/j.jweia.2007.01.011

Baker, C.J., Sterling, M., 2017. Modelling wind fields and debris flight in tornadoes. J. Wind Eng. Ind. Aerodyn. 168, 312–321. https://doi.org/10.1016/j.jweia.2017.06.017

Burgers, J.M., 1948. A Mathematical Model Illustrating the Theory of Turbulence, in: Von Mises, R., Von Kármán, T. (Eds.), Advances in Applied Mechanics. Elsevier, pp. 171–199. https://doi.org/10.1016/S0065-2156(08)70100-5

Chen, G., Lombardo, F.T., 2019. An analytical pattern-based method for estimation of a near-surface tornadic wind field. J. Wind Eng. Ind. Aerodyn. 194, 103999. https://doi.org/10.1016/j.jweia.2019.103999

Federal Emergency Management Agency (FEMA), National Institute of Standards and Technology (NIST), 2023. Design Guide for New Tornado Load Requirements in ASCE 7-22. https://tsapps.nist.gov/publication/get_pdf.cfm?pub_id=935883

Holmes, J. D., Baker, C.J., Tamura, Y., 2006. Tachikawa number: A proposal. J. Wind Eng. Ind. Aerodyn. 94, 41–47. https://doi.org/10.1016/j.jweia.2005.10.004

Holmes, J.D., Letchford, C.W., Lin, N., 2006. Investigations of plate-type windborne debris—Part II: Computed trajectories. J. Wind Eng. Ind. Aerodyn. 94, 21–39. https://doi.org/10.1016/j.jweia.2005.10.002

Huo, S., Hemida, H., Sterling, M., 2020. Numerical study of debris flight in a tornado-like vortex. J. Fluids Struct. 99, 103134. https://doi.org/10.1016/j.jfluidstructs.2020.103134

Jafari, M., Sarkar, P.P., 2020. Wind-induced response characteristics of a yawed and inclined cable in ABL wind: Experimental- and numerical-model based study. Eng. Struct. 214, 110681. https://doi.org/10.1016/j.engstruct.2020.110681

Kakimpa, B., Hargreaves, D.M., Owen, J.S., 2012a. An investigation of plate-type windborne debris flight using coupled CFD–RBD models. Part II: Free and constrained flight. J. Wind Eng. Ind. Aerodyn. 111, 104–116. https://doi.org/10.1016/j.jweia.2012.07.011

Kakimpa, B., Hargreaves, D.M., Owen, J.S., 2012b. An investigation of plate-type windborne debris flight using coupled CFD–RBD models. Part I: Model development and validation. J. Wind Eng. Ind. Aerodyn. 111, 95–103. https://doi.org/10.1016/j.jweia.2012.07.008

Kuligowski, E.D., Lombardo, F.T., Phan, L.T., Levitan, M.L., Jorgensen, D.P., 2014. Technical investigation of the May 22, 2011, tornado in Joplin, Missouri. Rep NIST NCSTAR 3.

Karstens, C.D., Gallus, W.A., Lee, B.D., Finley, C.A., 2013. Analysis of tornado-induced tree fall using aerial photography from the Joplin, Missouri, and tuscaloosa–birmingham, Alabama, tornadoes of 2011*. J. Appl. Meteorol. Climatol. 52, 1049–1068. https://doi.org/10.1175/JAMC-D-12-0206.1.

Lee, K.H., Rosowsky, D.V., 2005. Fragility assessment for roof sheathing failure in high wind regions. Engineering Structures 27, 857-868. https://doi.org/10.1016/j.engstruct.2004.12.017

Lin, N., Holmes, J.D., Letchford, C.W., 2007. Trajectories of Wind-Borne Debris in Horizontal Winds and Applications to Impact Testing. J. Struct. Eng. 133, 274–282. https://doi.org/10.1061/(ASCE)0733-9445(2007)133:2(274)

Lombardo, F.T., Roueche, D.B., Prevatt, D.O., 2015. Comparison of two methods of nearsurface wind speed estimation in the 22 May, 2011 Joplin, Missouri Tornado. J. Wind Eng. Ind. Aerodyn. 138, 87–97. https://doi.org/10.1016/j.jweia.2014.12.007.

Luo, J., Liang, D., Weiss, C., 2015. Reconstruction of a near-surface tornado wind field from observed building damage. Wind Struct. 20, 389–404. https://doi.org/10.12989/was.2015.20.3.389.

McDonald, J.R., 1976. Tornado-generated missiles and their effects. In: Proceedings of the Symposium on Tornadoes, pp. 331-348.

Miller, C.S., Kopp, G.A., Sills, D.M.L., Butt, D.G., 2024. Estimating Wind Speeds in Tornadoes Using Debris Trajectories of Large Compact Objects. Mon. Weather Rev. 152, 1859–1881. https://doi.org/10.1175/MWR-D-23-0251.1

NWS Analyze, F. and S.O., 2025, NWS Analyze, Forecast and Support Office; URL http://www.nws.noaa.gov/om/hazstats.shtml

Rankine, W.J.M., 1882. A Manual of Applied Physics. Man. Appl. Phys.

Redmann, G., Radbill, J., Marte, J., Dergarabedian, P., Fendell, F., 1978. Wind field and trajectory models for tornado-propelled objects. Electr. Power Res. Inst. Rep. EPRI NP.

Refan, M., Hangan, H., 2018. Near surface experimental exploration of tornado vortices. J. Wind Eng. Ind. Aerodyn. 175, 120–135. https://doi.org/10.1016/j.jweia.2018.01.042

Richards, P.J., Williams, N., Laing, B., McCarty, M., Pond, M., 2008. Numerical calculation of the three-dimensional motion of wind-borne debris. J. Wind Eng. Ind. Aerodyn. 96, 2188–2202. https://doi.org/10.1016/j.jweia.2008.02.060

Rott, N., 1958. On the viscous core of a line vortex. Z. Für Angew. Math. Phys. ZAMP 9, 543–553. https://doi.org/10.1007/BF02424773

Roueche, D.B., Prevatt, D.O., Haan, F.L., Datin, P.L., 2015. An estimate of tornado loads on a wood-frame building using database-assisted design methodology. J. Wind Eng. Ind. Aerodyn. 138, 27–35. https://doi.org/10.1016/j.jweia.2014.11.011.

Saini, D., Shafei, B., 2022. Flight characteristics of rod-shaped windborne debris objects in atmospheric boundary layer winds. J. Wind Eng. Ind. Aerodyn. 227, 105073.

Tachikawa, M., 1988. A method for estimating the distribution range of trajectories of wind-borne missiles. J. Wind Eng. Ind. Aerodyn. 29, 175–184. https://doi.org/10.1016/0167-6105(88)90156-0

Tachikawa, M., 1983. Trajectories of flat plates in uniform flow with application to wind-generated missiles. J. Wind Eng. Ind. Aerodyn. 14, 443–453. https://doi.org/10.1016/0167-6105(83)90045-4

Twisdale, L.A., Dunn, W.L., Davis, T.L., 1979. Tornado missile transport analysis. Nuclear Engineering and Design 51, 295-308. https://doi.org/10.1016/0029-5493(79)90096-7

Visscher, B.T., Kopp, G.A., 2007. Trajectories of roof sheathing panels under high winds. J. Wind Eng. Ind. Aerodyn. 95, 697–713. https://doi.org/10.1016/j.jweia.2007.01.003

Vladimirova, E., Gong, M., 2024. Advancements and Applications of Wood-Based Sandwich Panels in Modern Construction. Buildings 14, 2359. https://doi.org/10.3390/buildings14082359

Wills, J.A.B., Lee, B.E., Wyatt, T.A., 2002. A model of wind-borne debris damage. J. Wind Eng. Ind. Aerodyn., Bluff Body Aerodynamics and Applications 90, 555–565. https://doi.org/10.1016/S0167-6105(01)00197-0

Xu, F., Yin, L., Chen, W.-L., Duan, Z.-D., Xiao, Y.-Q., 2020. Numerical study on coupling flight characteristics of a plate-type windborne debris. J. Wind Eng. Ind. Aerodyn. 205, 104319. https://doi.org/10.1016/j.jweia.2020.104319

Zhao, J., Yan, G., Han, D., 2021. A review of approaches to simulate windborne debris dynamics in wind fields. J. Wind Eng. Ind. Aerodyn. 212, 104597. https://doi.org/10.1016/j.jweia.2021.104597